\documentclass[aps,amsmath,amssymb,pr,reprint,groupedaddress]{revtex4-1}

\usepackage{graphicx}
\usepackage{dcolumn}
\usepackage{bm}
\usepackage{color}
\usepackage{umoline}

\begin{document}

\title{
Pressure-induced recovery of Fourier's law
\\
in one dimensional momentum-conserving systems
}


\author{Dye SK Sato}
\email[]{dice@eri.u-tokyo.ac.jp}
\affiliation{
Earthquake Research Institute, University of Tokyo, 1-1-1 Yayoi, Bunkyo, Tokyo 113-0032, Japan
}


\date{\today}

\begin{abstract}
%
We report the two typical models of normal heat conduction in one dimensional momentum-conserving systems. They show the Arrhenius and the non-Arrhenius temperature dependence.
We construct the two corresponding phenomenologies,
transition-state theory of thermally activated dissociation and the pressure-induced crossover between two fixed points in fluctuating hydrodynamics.
Compressibility yields the {\it ballistic} fixed point, whose scaling is observed in FPU-$\beta$ lattices.
\end{abstract}


\maketitle

\section{Introduction}
The microscopic root of 
macroscopic linear irreversible processes~\cite{de2013non} have been closely studied
through the research
of heat conduction~\cite{lepri2003thermal,dhar2008heat,lepri2015heat}, where 
the processes are reduced to Fourier's law,
\begin{eqnarray}
J=-\kappa\nabla T ,\hspace{10pt}\kappa\propto N^0.
\end{eqnarray}
Here $J, \nabla T, \kappa$ and $N$ express 
 energy currents, temperature
gradients, 
heat conductivity and 
the size of systems. 
The early works examined the connection between the relation 
that 
temperature gradient becomes the thermodynamic force
($J\propto\nabla T$) and the nonintegrability
 corresponding to
 phonon scattering~\cite{rieder1967properties,casati1984one,prosen1992energy}.
Dimensionality
is considered as an intrinsic factor for
the intensive property of heat conductivity, e.g. {\it phonon localization}~\cite{matsuda1970localization,rubin1971abnormal,o1974heat} and {\it anomalous heat conduction}~\cite{narayan2002anomalous}. The verification is still progressing~\cite{dhar2001heat,lepri2003thermal,kundu2010heat,saito2010heat}.

Particularly, 
a universal breakdown of Fourier's law in one dimensional momentum-conserving systems is supported by theories~\cite{narayan2002anomalous,spohn2015fluctuating}, numerical simulations~\cite{lepri1997heat,hatano1999heat,savin2002heat,lepri2003thermal,mai2007equilibration} and experiments~\cite{chang2008breakdown}.
This systematic breakdown of the intensive property is called {\it anomalous heat conduction} and 
 heat conductivity in these systems grows 
with the power of $N$,
\begin{eqnarray}
\kappa\propto N^\alpha,\hspace{10pt} 0<\alpha< 1.
\end{eqnarray}
This universality is described by 
a semi-macroscopic continuum theory, {\it fluctuating hydrodynamic equations}~\cite{landau1987fluid} and 
the
recent research~\cite{grassberger2002heat,mendl2013dynamic,spohn2015fluctuating,das2014numerical} discovered its nontrivial relation with 
KPZ class,
a broad 
class of dynamical critical phenomena~\cite{kardar1986dynamic}.
The prediction of the theory and simulations clarified that
$\alpha$ takes the universal value~\cite{hatano1999heat,grassberger2002heat,narayan2002anomalous,mai2007equilibration,spohn2015fluctuating},
\begin{eqnarray}
\alpha = 1/3.
\end{eqnarray}
The recent research also has deeply studied the scaling property of fluctuations and revealed the undoubted evidence of the connection
that some scalings in these systems can be described by the KPZ equation
~\cite{das2014numerical,mendl2013dynamic,mendl2015current,spohn2015fluctuating,lepri2015heat}.
Anomalous
heat conduction forms 
the unique class connecting with 
KPZ class. 
Heat conductivity in one dimensional momentum-conserving systems is considered to show the divergence in the thermodynamic limit. 

The divergence of 
transport coefficients is considered as a general property of low dimensional momentum-conserving systems
and as the universal behavior of a fixed point in the fluctuating hydrodynamics 
~\cite{kawasaki1965logarithmic,pomeau1975time,forster1977large,narayan2002anomalous,mai2006universality}.
If 
systems do not
 conserve 
momentum, 
they do not
show such anomaly~\cite{hu2000heat,bricmont2007fourier,pereira2006normal}.
As the ordinary understanding of 
anomalous heat conduction, the correlation function of
spatially integrated energy currents $\hat J$ has the long time tails,
\begin{eqnarray}
\frac 1 N \langle\hat J(0)\hat J(t)\rangle^{eq}_{T,P} \sim t^{\alpha-1},
\end{eqnarray}
in $t_{m}\ll t <cN$ ($t_{m}$: microscopic characteristic time scale, $c$: sound velocity)~\cite{lepri1998anomalous,hatano1999heat,grassberger2002heat,narayan2002anomalous},
and it corresponds to the anomalous scaling. We used $\langle\rangle^{T,P}$ as the isothermal isobaric equilibrium average.
Equilibrium 
fluctuations of 
total energy currents
 $\hat J$ and 
heat conductivity are 
associated with 
a
Green Kubo formula~\cite{kubo2012statistical},
\begin{eqnarray}
\kappa(T;P)= \lim_{N\to\infty} \frac 1 {NT^2} \int ^\infty_0 dt \langle\hat J(0)\hat J(t)\rangle^{eq}_{T,P}.
\end{eqnarray}
Recent research~\cite{das2014numerical,spohn2015fluctuating} clarified that the correlation of the heat mode, one of the plane wave in perfect fluids contributes to the anomaly. 
The heat-mode autocorrelator $S_h(x,t)$ takes the asymptotic scaling form,
\begin{eqnarray}
S_h(x,t) \sim t^{-1/z}f_I(x/t^{1/z}) 
\end{eqnarray}
$f_I$ is the Levy distribution function. The exponent is another characterization of the fixed point,
\begin{eqnarray}
z=3/2.
\end{eqnarray}
Two exponents are connected with the dimension $d(\leq2)$ e.g. through the renormalization group approach~\cite{narayan2002anomalous}
 as ($z=1+d/2, \alpha =1-d/z(+0)$).
Even in the solids, the coefficient shows the divergence because of momentum conservation and dimensionality~\cite{mai2006universality,chang2008breakdown}.
The anomaly has been studied mostly 
with
the theory of fluids, but the mechanism is quite robust against such additional orders.

However, in spite of the completeness, 
the breakdown of this universality 
has been reported with recent detailed molecular dynamics
~\cite{gendelman2000normal,giardina2000finite,das2014role,zhong2011normal,chen2012breakdown,savin2014thermal,das2014heat,gendelman2014normal,wang2013validity,di2015anomalous,chen2014nonintegrability}.
The findings began with a model of poly-stable interaction potential energy, coupled rotator model
~\cite{gendelman2000normal,giardina2000finite}. 
This model 
shows the convergence of heat conductivity at large sizes and has been studied to settle
the unified sight between the anomaly and the 
{\it recovery of Fourier's law}
~\cite{das2014role}.
Following it, the subsequent research has reported the abundant systems of the recovery
~\cite{zhong2011normal,chen2012breakdown,savin2014thermal}.
The detailed investigation continues~\cite{das2014heat,gendelman2014normal,wang2013validity,di2015anomalous,chen2014nonintegrability}. 
Some cases~\cite{zhong2011normal,chen2012breakdown} were pointed out their finite size effects~\cite{das2014heat},
but as the worst for the universality class, another report of the research~\cite{chen2014nonintegrability}
revealed the diatomic hardcore systems can show the recovery in spite of the well-known feature as a paradigm of anomalous heat conduction.
On the other hand, some authors~\cite{gendelman2014normal} reported the possibility of the convergence caused by the quasi-stable {\it dissociation}. 
They also referred to the contributions of higher loop orders in fluctuating hydrodynamics, 
which was not studied in the previous analytical works.
In the kinetic theory, the first foundation of the anomaly~\cite{kawasaki1965logarithmic}, such effects come from the deviation of the interactions from 
the description of the stochastic two body interactions (Boltzmann equations). 
So we may need the careful study of such many-body interactions, though they are thought to be the source of the anomaly.
Actually, the quite recent research 
reported that the 
systems of multi-particle collisions 
showed the normal heat conduction~\cite{di2015anomalous}.
One dimensional momentum-conserving systems may form another class from the anomalous heat conduction. 
The unified view 
between the class of normal heat conduction
 and the ordinary anomalous class 
is far from established.

In this paper,
we report the two different possibilities of the origin, thermally activated dissociation~\cite{gendelman2014normal} and the increase of pressure.
We firstly present the two typical models of normal heat conduction in 
one dimensional momentum-conserving systems through the simulations of steady heat conduction. One shows the Arrhenius temperature dependence 
consistent with the previous work~\cite{gendelman2014normal}.
The other one shows the non-Arrhenius behavior. 
The latter is novel and suggests 
another mechanism of the 
convergence.
Next we show that the latter one would be related to pressure 
through the observations of equilibrium current fluctuations.
We try to explain the two mechanisms phenomenologically based on the transition-state theory of the dissociation formation and on the full fluctuating hydrodynamic equations.
In the latter analysis, the cutoff of the anomaly comes from the response to the pressure fluctuations. In the approach of mode coupling theory (MCT), 
one can see the convergence and a feature related to the breakdown of the hyperscaling between $\alpha$ and $z$.
The renormalization group (RG) 
result suggests the 
recovery
caused by the crossover between two fixed points.
Compressibility yields the novel {\it ballistic} fixed point,
\begin{eqnarray}
z=1.
\end{eqnarray}
We found this scaling can be reproduced
 by
the scaling discussions of the RG flows on full fluctuating hydrodynamic equations 
along the same procedure with \cite{narayan2002anomalous}.
The corresponding crossover is observed in the FPU-$\beta$ lattices.
Our results provide a picture of the recovery that the anomalous scaling can be cut off at $N_*$,
the characteristic size of the dissociation or that of the crossover between the two fixed points,
\begin{eqnarray}
\kappa\sim\left\{
\begin{array}{ll}
N  ^\alpha ,& N\ll N_*
\\
N_*^\alpha ,& N\gg N_* .
\end{array}
\right.
\end{eqnarray}

The construction of this paper is as follows;
Firstly, in {\it Settings}, we set the system Hamiltonian 
and the two typical interaction 
potentials,
and describe the details of two observations to measure 
the system-size dependence of
heat conductivity.
Next in
 {\it Results}, we report three results in the corresponding subsections:
the first 
one for steady heat conduction, 
the second one for equilibrium fluctuations of energy currents
and
the third 
one for analysis.
In 
{\it Discussions}, we discuss the correspondence of our results with the ordinary theories and with numerical simulations and the suggestion for experiments.
We report the 
inviscid-ballistic
scaling crossover of 
FPU-$\beta$ lattices in the same section.
\section{Settings}
\subsection{Settings of systems}
We study the one dimensional $N$ particle Hamiltonian systems of nearest-neighbor interactions,
\begin{eqnarray}
\mathcal H (\Gamma)
=
\sum_{i=1}^N \left[
\frac{p_i^2}{2m_i}+U(x_i)
\right]+
\sum_{i=1}^{N-1} V(x_{i+1,i}),
\end{eqnarray}
with appropriate boundary conditions.
$\Gamma:=(x_1,..,x_N,p_1,..,p_N)$ is the phase space coordinate.
With pinning-less monatomic ($m_i=m,U(x)=0$),
where the system should show the anomalous heat conduction, 
we 
set the two typical models of normal heat conduction.
One is for 
thermally activated recovery, 
Pure Repulsive-$\delta$ (PR-$\delta$) model,
\begin{eqnarray}
V(x)= \frac g\delta x^{-\delta}.
\end{eqnarray}
The other is for 
pressure-induced recovery,
 FPU-$\beta$ model,
\begin{eqnarray}
V(x)&=&\frac{K}{2} x^2+\frac \beta 4 x^4.
\end{eqnarray}
These two models show the different forms of temperature dependence on heat conductivity with each other.
The former one suggests the non-continuum 
mechanism as already reported in~\cite{gendelman2014normal},
and the latter one 
provides an inconsistent example with their class.

Pure Repulsive-$\delta$ is described here.
The mechanical parameters are ($m, g, \delta, d, e$). 
$d$ is the average distance of particles given by the boundary conditions and $e$ is the energy scale per particle
given by the initial conditions or by attached thermostats.
The free parameters are ($\delta,V_*:=V(d)/e$). $x_i$ expresses the $i$-th particle position. 
Two limits $V_*\to 0$(dilute high-energy), $\infty$ (dense low-energy) with finite $\delta$ correspond to integrable limits 
(of hardcore particles and of harmonic chains).
This
 system shows the strong nonlinearity at $V_*\sim 1$. $\delta$ characterizes the interaction decay ($\delta\to 0$: log interactions, $\delta\to\infty$: delta interactions).
It 
has been already
reported 
this 
system 
should
show the asymptotic 
convergence of heat conductivity 
at $\delta=1,6,12$~\cite{savin2014thermal}.
Now we take the unit $(m, g) \to1$ and choose the parameters as ($\delta=6,V_*\sim1$) to get the strong nonlinearity
\footnote{In the sense that the maximum Lyapunov exponent is large 
compared with dilute case ($d>1$) and the Lyapunov spectra shows the same shape with high energy FPU-$\beta$, which correspond to the case that the Hessian of Hamiltonian is able to be replaced by the random matrix~\cite{newman1986distribution} on the surface spanned by conserved quantities.}.

FPU-$\beta$ has following mechanical properties.
($m, K, \beta, l, e$) are the mechanical parameters
of this system. 
$l$ is the averaged compression given by the boundary conditions. $e$ is the same with the PR-$\delta$ case.
$x_i$ expresses the deviation from the equilibrium point
under the free boundary. An appropriate center uniquely determines it.
The compression parameter $l$ expresses the mismatch
 between 
the equilibrium position of force under the chosen boundary and that under the free boundary.
We take the unit $(m, K, \beta) \to 1$.
The left 
free parameters are $(l, e)$ here.
 $e$ decides the nonlinearity of the system
($e\to 0$: harmonic, $e\to \infty$: strong nonlinear),
and $l$ corresponds to the pressure.
The heat conduction at ($e\gg1,|l|\ll1$) is well studied~\cite{lepri2003thermal}, and it shows the fluctuating hydrodynamic (FH) scaling at large $N$~\cite{lepri1997heat,mai2007equilibration}.
Our 
simulations suggest 
the compression would be intrinsic for the 
recovery of Fourier's law.
For the notation, we introduce the isodense-potential (coordinate trans. $x_n\to y_n:=x_n+nl$)
\begin{eqnarray}
V(x)=\frac{K}{2} x^2+\frac\beta 4 x^4 
\hspace{5pt}
\to
\hspace{5pt}
\frac{K}{2} (x-l)^2+\frac\beta 4 (x-l)^4. 
\end{eqnarray}
We call it compressed FPU-$\beta$ (c-FPU-$\beta$) here. 
The isodense condition is tricky in FPU-$\beta$ 
($l$-compressed fixed: $V(x_1)+(V(-(N+1)l-x_N)$,
$l$-compressed periodic:$V(x_1-(x_N+Nl))$),
but easy in c-FPU-$\beta$
(fixed: $V(x_1)+V(-x_N)$, periodic: $V(x_1-x_N)$).
We study its $l$ dependence in highly nonlinear regime ($e\gg1$), 
where 
the system shows the FH anomaly at $l=0$, zero pressure~\cite{lepri1997heat,mai2007equilibration}.
\subsection{Settings of experiments}
We studied 
two observables to investigate the system size dependence of heat conductivity.
One is the heat conductivity defined by 
heat currents 
and 
temperature profiles
under 
the states of 
steady heat conduction 
maintained by stochastic reservoirs.
The other 
is 
power spectra of 
total energy currents 
under the isolated equilibrium  conditions with periodic boundaries.
We note that we studied 
heat conductivity of the former case
under the globally far from equilibrium conditions 
where the temperature difference of the attached reservoirs is comparable with the average temperature of them.
If the system recovers the 
intensive property of heat conductivity, 
this observation lets us see the temperature dependence of the heat conductivity under the isobaric conditions.
The temperature dependence connects to the possible mechanisms of the recovery. 
As an example, the dissociation class shows the Arrhenius temperature dependence of heat conductivity~\cite{gendelman2014normal}.

The details of 
time integrations and of reservoirs are as follows.
We use the fourth order symplectic integrator with 
time steps $\Delta t=0.0025$ in PR-6 and with $\Delta t =0.02$ in c-FPU-$\beta$.
We also checked the independence of 
our 
results on time 
steps by the changing 
$\Delta t\to \Delta t/4$ in some experiments.
For PR-6, we set the thermal wall boundaries~\cite{hatano1999heat} at the both edges ($x=0,Nd$).
For c-FPU-$\beta$, we introduce the potential boundaries ($V(x_1)+V(-x_N)$) and 
attach the boundary reservoirs of Langevin type (viscosity: $\gamma$) 
to the $10+10$ particles at the corresponding edges. 
The algorithm of the attachment is 
geometric 
Langevin type~\cite{bou2010long}.
The reservoir parameters are chosen as $(T_L,T_R,d)=(4,1,3/4)$ for PR-6 and
$(T_L,T_R,\gamma)=(50,30 (10),2)$
for c-FPU-$\beta$.
We make the initial conditions with the attachment of Langevin reservoirs 
in equilibrium cases. 

\subsubsection{Definitions of temperature and heat conductivity}
\begin{figure}[tbp]
  \includegraphics[width=80mm]{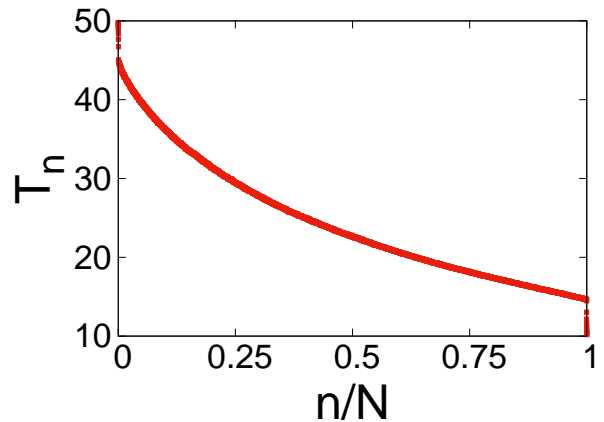}
 \caption{
The 
temperature
profile of c-FPU-$\beta$ at 
$(T_L,T_R,l,N)=(50,10,4,2^{15})$.
One can see the gaps and the varying gradient.
 }
 \label{fig:one}
\end{figure}

We also note about the arbitrariness of the defined heat conductivity.
One can define 
energy currents 
and 
temperature profiles 
based on the particle index or on the field coordinate~\cite{dhar2008heat}. This is 
a property of the lattice systems.
This arbitrariness makes following confusing situations under the globally far from equilibrium conditions.
The density heterogeneity causes the two different temperature gradients,
although the steady state values of the energy currents show the quantitative correspondence
($\langle J_{field}\rangle=\langle J_{particle}\rangle$ for PR-6).
Then the values of heat conductivity defined by the ratios are different. 
This difference remains 
even though the system is of nearly local equilibrium, 
because the system satisfies the isobaric (not isodense) conditions under the steady heat conduction.
In our PR-6 case, however, we observe the density difference takes a rather small value, roughly 10 percent of initial packing density $d^{-1}$.
This is negligible for the discussion of the anomaly.
In compressed FPU-$\beta$ case there is no arbitrariness, because one can take the lattice constant as large as one wants~\cite{mai2007equilibration}.
We take the particle type observables,
e.g. $J_{n+1, n}:=\frac{p_{n+1}+p_n}2\left(-\frac{\partial V(x_{n+1,n})}{\partial x_{n+1,n
}}\right)$ with energy density $e_n:=p_n^2/(2m)+(V(x_{n+1,n})+V(x_{n,n-1}))/2$, through this paper.
One can also define the stress along particle coordinates 
$\sigma_{n+1,n} = -\frac{\partial V(x_{n+1,n})}{\partial x_{n+1,n}}$.
It gives the pressure of the steady state (in the sense of particle coordinates) conforming to the virial theorem.

\section{Results}
\subsection{Result of Nonequilibrium Simulation}
Now we observe the kinetic temperature, 
which corresponds to the thermodynamic one under the local equilibrium condition
 (, $(T_{n+1}-T_n)/T_n\ll1$ for the finite interparticle distance).
Here, the kinetic temperature is defined as 
\begin{eqnarray}
T_n:=\left\langle \frac{p_n^2}{m_n} \right\rangle.
\end{eqnarray}
We use the bracket $\langle\rangle$ to express the long time average. 

Figure \ref{fig:one} is the temperature profile of c-FPU-$\beta$ under the steady heat conduction at 
$(T_L,T_R,l,N)=(50,10,4,2^{15})$
and
shows the gaps at the edges and the gradient varying at the bulk.
As $N$ increases,
the gaps are slowly decreasing and the varying gradient remains 
\footnote{
This is a prefetch but one cannot see the gradient increasing at the edges 
in this case, which is a characteristic feature of anomalous systems~\cite{van2012exact}.}.
The situation of the PR-6 case is the same with 
it. 
We consider the points.
Firstly, it is necessary to remove the boundary gap, which can show the 
convergence 
of pretense~\cite{hatano1999heat}, 
for the accurate observation of the system size dependence.
Secondly, the spatially varying gradient means the spatially varying heat conductivity.

Then we define 
heat conductivity in two ways to discuss its system size dependence and its spatial variance.
$J_{n+1,n}$ is independent of the index in the average of the steady state if the $n,n+1$th particles have no interaction with reservoirs. So we simply call it $J$.
Firstly, to study the system size dependence, 
we introduce the bulk averaged gradient $\overline{ \nabla_n T_n}$, which is given by the linear fitting of $T_n$ in 
$1/4\leq n/N\leq 3/4$, and define the following heat conductivity,
\begin{eqnarray}
\bar \kappa(N) := \langle J\rangle/\overline {\nabla T}.
\end{eqnarray}
Here we call it bulk heat conductivity.
We discuss the anomaly of the bulk through this observable under an assumption that the effect of the discontinuities is negligible. 
This assumption is valid in nonintegrable large systems. 
These gaps occur within the reservoir-attached particles at the edges, so they come from purely interfacial resistance.
Secondly, 
to study the spatial variance,
we introduce the segmented averaged gradient 
$\widehat{\nabla_n T_n}$, which is given by the linear fitting of
$T_n$ in a sufficiently narrow temperature range sufficiently slowly varying,
and define the following heat conductivity
 with the assumption that 
$\widehat{\nabla_n T_n}$ 
is a function of the average temperature of each segment and $N$, 
\begin{eqnarray}
\kappa(T;N) := \langle J\rangle/\widehat{\nabla T}.
\end{eqnarray}
Now we call it local heat conductivity. 
We took two ways to prepare the data of such a temperature range. 
One is that we equally divide the data 16 (or 32) and cut out $T_n$ of the reservoir-attached particles. 
The other is that we set ($T, \Delta\ll T$) and extract $T_n$ satisfying the condition $|T_n-T|<\Delta$.
Our assumption of local heat conductivity is valid in the systems of normal heat conduction.
Heat conductivity of normal heat conduction depends only on the local thermodynamic quantities,
although that of anomalous heat conduction depends also on the positions and on the system sizes.
Then, this observable lets us know the temperature dependence of the heat conductivity under the isobaric conditions of the pressure $P$ given by the boundary conditions, if the system recovers the 
intensive property of heat conductivity. 
We study the asymptotic master curve of the 
intensive heat conductivity. 
We can observe such temperature dependence at once
under the globally far from equilibrium conditions. 

\begin{figure*}[tbp]
   \includegraphics[width=165mm]{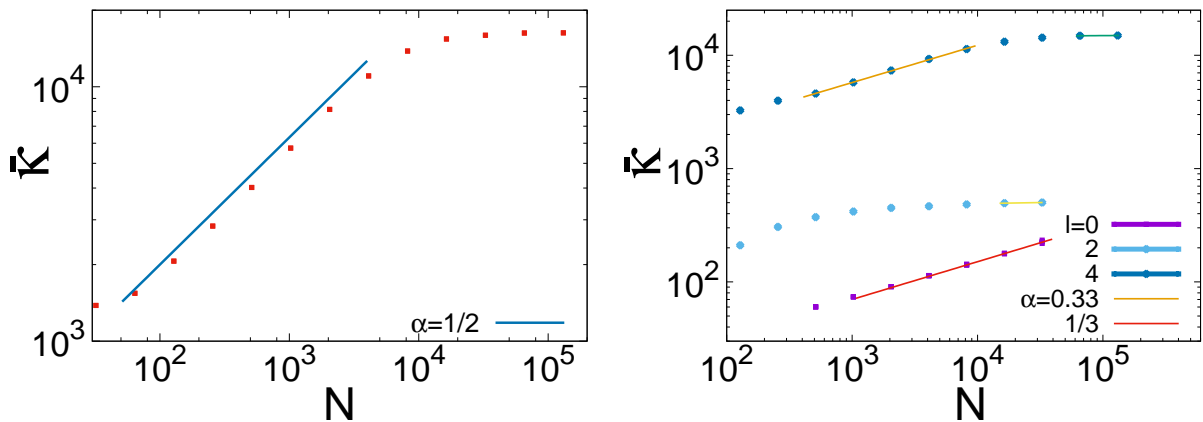}
  \caption{
Bulk heat conductivity (Left: PR-6, Right: c-FPU-$\beta$).
Parameters are $(T_L,T_R,d)=(4,1,3/4)$ for PR-6 
and 
$(T_L,T_R,\gamma,l)=(50,30,2,4)$
 for c-FPU-$\beta$.
They show the convergence except c-FPU-$\beta$ at $l=0$.
The transient exponents are $1/2$ in PR-6 and 
$0.33$ 
in c-FPU-$\beta$ at $l=4$.
For accuracy, we check the power of the last two points in c-FPU-$\beta$ at $l=2,4$
as an indicator of the convergence.
The powers are 
$0.018(l=2)$ and $0.009(l=4)$, 
which show the quantitative evidence of the convergence.
  }
  \label{fig:two}
\end{figure*}
\subsubsection{System size dependence of bulk heat conductivity}
Figure \ref{fig:two} is the result 
of the bulk heat conductivity $\overline\kappa$
 and shows the 
convergence 
in some ways.
$\overline \kappa$ of PR-6 shows the transient anomalous scaling $\alpha=0.498\pm0.005$ in $10^2\alt N\alt 10^4$ (the exponent of which is fitted in $2^{7}\leq N\leq2^{12}$ )
and the recovery of normal heat conduction without showing the FH scaling $\alpha=1/3$.
$\overline \kappa$ of c-FPU-$\beta$ is somewhat strange. 
It shows the FH anomalous scaling $\alpha=1/3$ at $l=0$, 
consistent with previous reports~\cite{lepri1997heat,mai2007equilibration}, but it 
shows the 
convergence 
at $l=2,4$.
Particularly, at $l=4$, it shows the anomalous scaling 
$\alpha=0.330\pm0.006$
 in
 the transient $N$-dependence (fitted in 
$2^{9}\leq N\leq2^{13}$
), which is almost the FH exponent 1/3.
It suggests that the 
convergence
should occur after the FH scaling, in the continuum limit.

$\overline \kappa$ of PR-6 shows the divergence with the exponent $\alpha\simeq 1/2$
in 
$10^2\alt N\alt10^4$
and the convergence at larger sizes.
This convergence 
is consistent with~\cite{gendelman2014normal}.
Its exponent $\alpha\simeq1/2$ is larger than FH scaling $\alpha=1/3$ and just the second universality exponent~\cite{van2012exact,hurtado2015violation}.
The intensive behavior remains even at $N\sim10^5$.
As recently reported, diatomic hardcore shows the convergence if their chosen parameters are nearly integrable~\cite{chen2014nonintegrability}. 
Some nearly integrable systems may show significant slow-down to obey the prediction of the continuum theory like in the FPU-problem~\cite{spohn2015fluctuating}. 
However, our choice of parameters provides the strong nonlinearity, 
although PR-6 has integrable in some limits
(See {\it Settings}). We need another explanation of the convergence.
It might generate the ballistic transport of the hardcore 
by the local thermal activation 
to cause this result. 
This mechanism is consistent with the dissociation-induced 
recovery
 proposed in~\cite{gendelman2014normal}.

$\overline \kappa$ of c-FPU-$\beta$ shows the anomaly at $l=0$ but 
the converging behavior at $l=2,4$ and we observe the transient anomaly at $l=4$ 
where the exponent is the FH $\alpha\simeq1/3$.
Firstly, the result at $l=0$ is consistent with the ordinary understanding of 
anomalous heat conduction~\cite{lepri1997heat,mai2007equilibration}.
The difference of the trend from the previous results 
at 
$l=0$ with small sizes
is caused by the difference of reservoirs and 
by 
the definitions of the temperature gradient. In our case, the number of the attached particles is slightly larger,
so the boundary resistance is rather small, and the direct measurement of gradients reduces the effect of temperature gaps
to the observable. 
Secondly, $\overline \kappa$ at $l=2$ changes the $N$-dependence completely from the case at $l=0$ and 
 shows the converging behavior throughout the observation. This is inconsistent with 
ordinary predictions~\cite{narayan2002anomalous}, then
we calculated $\overline \kappa$  until 
 $N>10^4$ 
but the results were the same.
Here we introduce the power measured from the last two points as the indicator of convergence.
The power is 
$\alpha\simeq0.018$ 
at $l=2$, almost 20 times smaller than the predicted exponents.
The increase goes to be saturated.
At the last, $l=4$ is the delicate case. We got the transient exponent 
$\alpha\simeq1/3$
in 
$10^2\alt N\alt10^4$
, which is consistent with the FH theory, 
but $\overline \kappa$ shows another trend at large $N$.
There is a plateau at larger sizes.
We checked the power from the last two points and got 
$0.009$.
It is 
the convergence.
The weak exponents in FPU-$\alpha\beta$ at some parameters 
are 
already observed in~\cite{savin2014thermal}, but the converging behavior like our case was not reported in their 
studies. 
We can see the transient exponent smaller than 0.1 in $l=2$ case, so it may correspond to their reports of such weak exponents. 
The difference would be caused by the temperature or by finite size effects. 
We chose the high temperature ($T\geq10$), highly nonlinear, and $N>10^4$, 
but their choice was low $T(\leq 0.1)$, weakly nonlinear, and $N<10^4$.
Actually, the long time tail is sensitive to the energy. 
The result in~\cite{di2015anomalous} showed the normal-anomalous crossover around $e\sim0.1$ in FPU-$\alpha\beta$ (at $\alpha=0.1$) of the same unit with us.
Repeatedly we note the high-temperature of our system.
It is also hard to suppose the possibility of dissociation 
because of the system property. This is directly checked by the following results 
of the temperature dependence in the local heat conductivity.

One can 
doubt whether our results are transient 
plateaus.
The flattening before the FH scaling is actually reported in the previous research~\cite{das2014heat}.
Our case 
of FPU-$\beta$
also has such a flattening at 
$l=2,4$ 
in $N\alt 10^3$.
However, one can see the convergence 
which is after the FH scaling at $l=4$.
This is not included in their 
studies.
next we study the local heat conductivity to avoid the 
rash conclusion,
although our results
of bulk heat conductivity
suggested the 
recovery of
the 
intensive property accompanying the 
increase of pressure. 

\subsubsection{Temperature dependence of \\intensive local heat conductivity}
\begin{figure*}[tbp]
   \includegraphics[width=165mm]{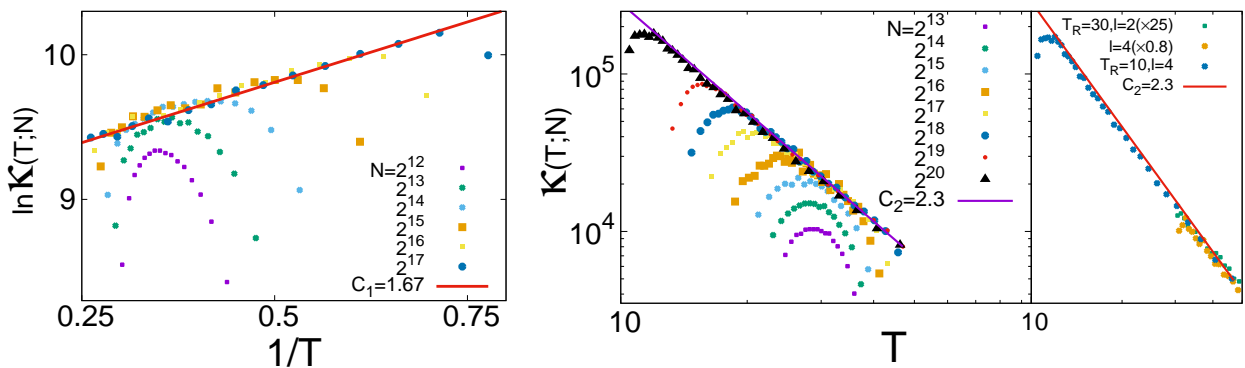}
  \caption{
Local heat conductivity 
(Left: PR-6, Center and Right: c-FPU-$\beta$).
Parameters are $(T_L,T_R,d)=(4,1,3/4)$ for PR-6 
and 
$(T_L,T_R,l)=(50,10,4)$
 for c-FPU-$\beta$ 
in the center panel.
The right panel shows the values of FPU-$\beta$ at the largest size of our observations with multiplication of normalization factors. They can be fitted with a unique curve approximately.
One can 
observe the two different types of temperature dependence:
the Arrhenius type 
connected with 
the thermally activated 
recovery in PR-6
and the non-Arrhenius type 
in c-FPU-$\beta$, suggesting another 
process from thermal activation.
  }
  \label{fig:three}
\end{figure*}

Figure \ref{fig:three} is the result of local heat conductivity $\kappa$ 
and shows the two asymptotic master curves $\kappa(T;\infty)$
accompanying 
the recovery of normal heat conduction.
Here we call the suggesting master curve $\kappa(T;\infty)$. Although
the secured convergence of $\kappa(T;N)\to\kappa(T;\infty)$ 
in 
the thermodynamic limit is still controversial, these $N$-independent curves are the strong evidence of the recovery.
Apparent $N$-dependence of $\kappa$ means the breakdown of our assumption
that the local heat conductivity is described by the local $(T,P)$, 
and one can see such $N$-dependence only at small sizes.
From the master curves, one can see the Arrhenius $T$-dependence in PR-6 which corresponds to the thermal activation of dissociation~\cite{gendelman2014normal} and the non-Arrhenius one in c-FPU-$\beta$,
which corresponds to 
a different mechanism
from thermal activation.

$\kappa$ of PR-6 shows the recovery of the intensive property  
and 
the Arrhenius temperature dependence,
\begin{eqnarray}
\kappa(T,\infty)\sim e^{C_1/T}.
\end{eqnarray}
The recovery began from the bulk.
$C_1$ is estimated as $C_1\simeq 1.67\pm0.04$ from our fitting.
It is already reported that soft rod systems also show the Arrhenius type heat conductivity given by the GK formula~\cite{gendelman2014normal},
and they claimed the effective vacancy, dissociation should cause the normal heat conductivity.
Such a mechanism would be also realized in this case.
The local dissociation is naturally understood as the high energy property 
in this system (integrable hardcore).

$\kappa(T;N)$ of compressed FPU-$\beta$ 
at $(l,T_R)=(4,10)$
recovers
 the intensive property from the hotter 
side of the 
bulk
and converges to the power-law 
curve,
\begin{eqnarray}
\kappa(T;\infty) \sim T^{-C_2}.
\end{eqnarray}
$C_2$ is estimated as $C_2\simeq 2.30\pm0.01$ from our fitting. 
We note the replacement of $T_R$ ($30\to10$) is done to obtain the sufficiently 
wide temperature range. The result looks robust against the replacement. 
Figure \ref{fig:three} (Right)
shows the values of FPU-$\beta$ at the largest size of our observations 
($(T_R,l)=(30,2),(30,4),(10,4)$, corresponding size $N=2^{15},2^{17},2^{20}$) 
with multiplication 
of normalization factors ($\times25,0.8,1$ respectively). They can be roughly fitted with a unique curve.
The non-Arrhenius $T$-dependence suggests another mechanism from some thermal activation.
Taking into consideration
the high $T$ property of this system, the result of bulk heat conductivity and this non-Arrhenius $T$-dependence,
one can expect that this 
recovery of the intensive property
may be understood with some continuum theory.
We also observed the different exponent from already reported $\kappa\sim T^{1/4}$ scaling of zero pressure FPU-$\beta$~\cite{aoki2001fermi}.
The cause may be provided from the 
convergence
of the heat conductivity or the pressure dependence of the heat conductivity
because of the effect that the density fields of FPU-$\beta$ becomes spatially non-uniform to satisfy the constraint of uniform $(P,J)$ conditions except $P=0$.
We also observed almost the same scaling at $l=2$,
 then it 
may come from the normal-anomalous crossover.
More accurate discussions on this change of the $T$-dependence across the crossover need other experiments from ours.
We firstly took the assumption that the global coupling is negligible. 
This assumption cannot be applied in anomalous heat conduction.
Our interest is the difference of the temperature dependence from the already reported Arrhenius one and it was shown, so
we do not go into the detailed discussion of this difference here.

\subsection{Result of Equilibrium Simulation}
It was suggested through the observations of steady heat conduction 
that there should be a thermally activated 
recovery of the heat conduction and 
that of continuums.
Particularly, the latter case 
showed a striking negative example for 
ordinary fluctuating hydrodynamic predictions.
Then we also study the 
convergence of heat conductivity
 through the observations of 
total energy current 
power spectra under the isolated equilibrium periodic boundary conditions
to check our problematic result. 
Here we assume the equivalence of ensembles and change the boundary conditions from the isobaric isothermal $\langle\rangle^{eq}_{T,P}$ to the isothermal isodense $\langle\rangle^{eq}_{T,l}$.
We studied its power spectra $|\hat J(\omega)|^2$, 
the Fourier components of the correlations $\langle\hat J(0)\hat J(t)\rangle^{eq}_{T,l}$, and their $l$-dependence.

\begin{figure*}[tbp]
  \includegraphics[width=160mm]{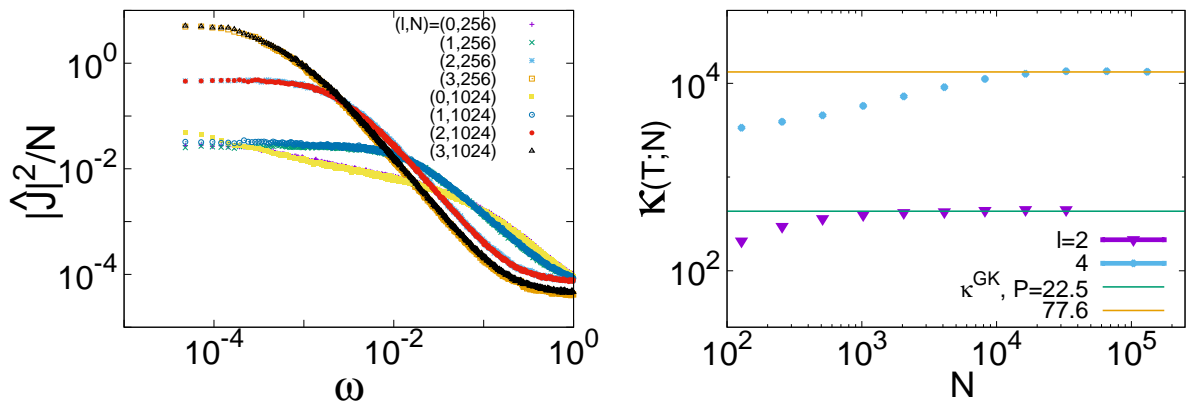}
 \caption{
Left: Normalized power spectra of the total energy currents $|\hat J(\omega)|^2/N$ 
of c-FPU-$\beta$ at $T=30$ in $10^{-5}\alt\omega\alt1 $.
Right: Comparison between the local heat conductivity at the averaged temperature $\bar T$ of steady heat conduction and the corresponding value given by the Green Kubo formula.
One can see the anomalous scaling $|\hat J(\omega)|/N\sim \omega ^\alpha$
at small $\omega$ of $l=0$, but Lorentzian$|\hat J(\omega) |^2/N\sim a/(1+(b\omega)^2)$ recovers with $l$ increasing ($l=2,3$).
There is no system size dependence at $l=2,3$. 
The trends of (a,b) that (a,b) increase with $l$
are consistent with Figure \ref{fig:two}.
Pressure 
looks to suppress the long time tail.
The comparison of heat conductivity (Right) shows the consistent convergence of heat conductivity in highly compressed FPU-$\beta$.
 }
 \label{fig:four}
\end{figure*}

We beforehand discuss the validity of such equivalence. It is necessary to characterize the heat conductivity by a single value of temperature $T$
that the temperature difference
is sufficiently small in the correlation length of the energy current, 
but such a length does not exist in the systems of anomalous heat conduction~\cite{pomeau1975time,das2014numerical}.
The quantitative correspondence of the heat conductivity in globally far from equilibrium with 
the correlation function in the isolated periodic boundary is actually a delicate problem~\cite{casati2003anomalous,das2014heat}.
Nevertheless, one can observe the long time tails of the currents
under the isolated periodic boundary conditions
and 
these scalings of the anomaly show the correspondence even with such changes of boundaries (see the $l=0$ case of Figure \ref{fig:four}),
furthermore, one can see the correspondence well because of the absence of boundary perturbations~\cite{lepri1998anomalous,hatano1999heat}.

Figure \ref{fig:four} 
(Left) 
shows
normalized 
power spectra
of total energy currents
 $|\hat J(\omega)|^2/N$ in compressed FPU-$\beta$ at $T=30$.
The anomalous scaling $|\hat J(\omega)|^2/N\sim\omega ^{-\alpha},\alpha=1/3$ is observed in the $l=0$ case of zero pressure. 
The strong exponent observed at high frequencies corresponds to the next FH exponent $4/3$.
However,
$l=1$ case shows no FH scaling at low frequencies 
although one can see the next FH $4/3$ tails at high frequencies.
Then it asserts the FH tail in low frequencies is suppressed by the compression. 
The scaling changes to the Lorentzian at highly compressed cases ($l=2,3$) and
one can see the clear suppression of the anomaly at low frequencies.
The trend in the Lorentzian of the height and that of the time scale are consistent with our experiment of steady state heat conduction that they increase with 
compression.
We checked the $N$-convergence of the Lorentzian
and the results at $N=256,1024$ were the same. 
In the ordinary understanding,
if the {\it ad hoc} cutoff of the anomaly $\omega>2\pi c/N$ is valid also under this condition,
the spectra at $l=1,2,3$ should show the long time tails 
at 
the same area with $l=0$ case, except the improbable explosion of $t_{m}$ with $l$-increasing, because the sound velocity is only doubled through the $l$-increasing ($l=0\to3$ at $T=30$).
The result that there is only the 4/3 exponent (no 1/3) at $l$ = 1 and the Lorentzian recovery at $l = 2,3$ suggests that the compression
(pressure)
should induce the suppression.

This result of equilibrium correlations should show the quantitative correspondence with our result of steady heat conduction if the convergence is not a fake. 
The Green Kubo formula requires that a variable given by the equilibrium correlations
$N^{-1}\int ^\infty_0 dt\langle\hat J(0)\hat J(t)\rangle^{eq}_{T,P}$ and heat conductivity of the system $\kappa_*(T,P,N)$ take a same value, that is 
\begin{eqnarray}
\lim_{N\to\infty}\left(
\frac 1 N \int ^\infty_0 dt\langle\hat J(0)\hat J(t)\rangle^{eq}_{T,P} -\kappa_*(T,P,N)
\right)=0.
\hspace{16pt}
\end{eqnarray}
In general, we should include the other orders of the system and in such a case,
one can translate $(T,P)$ into the whole intensive variables 
conjugate with the conserved fields of the system except $N$. 
We can expect FPU-$\beta $ should not have such additional orders because of the consistency between equilibrium correlations of FPU-$\beta$ and the prediction of fluctuating hydrodynamics for simple fluids~\cite{das2014numerical}. 
Furthermore, if the system shows the normal heat conduction,
the above relation becomes more tractable. 
Firstly, energy currents in the systems of normal heat conduction have the finite correlation length, 
so heat conductivity is not modified under the change of boundaries if the change does not vary $(T,P,N)$. 
It means 
\begin{eqnarray}
\frac 1 N \int ^\infty_0 dt\langle\hat J(0)\hat J(t)\rangle^{eq}_{T,P}
=
\frac 1 N \int ^\infty_0 dt\langle\hat J(0)\hat J(t)\rangle^{eq}_{T,l_*(T,P)}.
\nonumber\\
\end{eqnarray}
We defined $l_*(T,P)$ as the compression that gives the pressure $P$ under the temperature $T$ (and $N$). 
Also, in the system of normal heat conduction, heat conductivity is independent of the observational condition if $(T,P)$ take the same values. 
It is another consequence of the locality guaranteed in the systems of normal heat conduction. 
Then, from the above relations, the question about the consistency between 
the variable given by the equilibrium correlations, 
\begin{eqnarray}
\kappa^{GK}(T,l,N) := 
\frac 1 N \int ^\infty_0 dt\langle\hat J(0)\hat J(t)\rangle^{eq}_{T,l},
\end{eqnarray}
and our local heat conductivity $\kappa(T,N)$
gives us the hint whether the numerical convergence captures the behavior in the thermodynamic limit. 
For this investigation, it is enough to check the relation,
\begin{eqnarray}
\lim_{N\to\infty}\left(
\kappa^{GK}(T,l_*,N)-\kappa(T,N)
\right)
=0.
\end{eqnarray}
We note the Green Kubo relation holds even at finite sizes if the observational condition and the size are the same, which can be proved with fluctuation theorem~\cite{seifert2012stochastic} (see {\it Appendix}), but this is not guaranteed in different sizes or with different boundaries. 
So, this consistency becomes strong evidence of the convergence in the thermodynamic limit, though we cannot access the limit in rigorous sense.

Figure \ref{fig:four} (Right) shows the consistency between $\kappa^{GK}(T,l_*,N)$ of equilibrium simulations and $\kappa(T,N)$ of steady heat conduction at the averaged temperature of reservoirs $T=\bar T(:=(T_L+T_R)/2)$. 
As subtle but significant notation, $l_*$ and $l$ of nonequilibrium simulations of far from equilibrium conditions should not take the same value in general, because local compression of nonequilibrium simulations takes a different value from $l$. Homogeneity of pressure under the simulation is satisfied for the steady condition, so the choice of pressure avoids meaningless confusion. 
The values of pressure in nonequilibrium simulations are 
$P\simeq 22.5 (l=2),77.6 (l=4)$ with Langevin reservoirs ($T_L=30,T_R=50,\bar T=40,\gamma=2,N=2^{15}$, attached particle number 10+10) at the edges.
The values of pressure were robust against the size in our case. 
We measured $\kappa^{GK}$ at the rather small sizes  ($N=1024 (l=2),256(l=4)$) 
on receiving the $N$-independence as seen in the left panel of Figure \ref{fig:four}.
Both cases of different compression $l=2,4$ show the quantitative correspondence with Green Kubo values at large sizes. 
The transient exponent at $l=4$ is consistent with the corresponding bulk heat conductivity.
This system shows
 the numerical consistency of two variables under the different conditions (of sizes and of boundaries). It strongly asserts the normal heat conduction which remains at the larger sizes.

\subsection{Result of Analysis}
Our simulations showed two different types of temperature dependence on intensive heat conductivity. 
They suggest two mechanisms in the recovery of Fourier's law. 
The Arrhenius form connects to a thermally activated inhibitor and the non-Arrhenius one suggests a continuum mechanism different from the former. 
Here, we try to explain these processes phenomenologically.

\subsubsection{Transition-state theory of \\thermally activated dissociation}

Here we 
estimate 
the coefficient of the Arrhenius 
$T$-dependence in PR-6 along the 
scenario 
of thermally activated dissociation proposed 
in~\cite{gendelman2014normal}.
In PR-6, 
there is no FH scaling before the 
convergence 
and  the temperature dependence is 
the Arrhenius 
form. 
This result suggests the possibility that some non-continuum effect would suppress the anomaly.
According to them, this 
convergence 
is caused by the effective vacancy (dissociation) formation. Now we assume it as an effective point defect
and estimate the effective activation energy to compare with our numerical result.

If one can define the maximum energy state on each vacancy formation path, according to the transition-state theory,
the vacancy formation ratio $P_*\sim \exp(-E_*/T)$ can be decided by the minimum of their energy cost~\cite{eyring1935activated},
i.e. $E_*$ is the energy difference between the maximum energy and the 
minimum
 energy on the path to generate a defect with the 
smallest
energy cost.
In this case, the path corresponds to the quasi-static path, then $E_*$ corresponds to the minimum energy benefit to make a vacancy at the bulk.
Concretely, as the minimum work to make a vacancy of the characteristic length $d$ at the bulk including 
$N$ 
particles, one can estimate $E_*$ as 
\begin{eqnarray}
E_*&\sim&
N
\left(V\left(d\left(1-\frac 1 {N}\right)\right)-V(d)\right)-V(d)
\\
&\simeq& -\frac{\partial V(d)}{\partial d} d-V(d).
\end{eqnarray}
Then, the minimum length which has a vacancy in average would be
\begin{eqnarray}
N_*\sim P_*^{-1}\sim e^{E_*/T}.
\end{eqnarray}
Then, assuming the scaling 
$\kappa\sim$ 
$N^\alpha$ 
is saturated at $N_*$, the temperature dependence of $\kappa$ is given by
\begin{eqnarray}
\kappa \sim 
N_*^\alpha
\sim
e^{\alpha E_*/T}
.
\end{eqnarray}
We compare the estimation with the numerical value: $C_1\simeq 1.7\sim 2$.
We derive the value: $C_1\sim 2.341\sim 2$ from the above discussion
with $\alpha\simeq1/2$ of the simulation.
The mechanism of the 
recovery 
caused by the thermally activated point defect 
 gives an approximate picture to this 
convergence, 
even though the dissociation is not well defined in this system.
One can take 
elastically colliding systems like in~\cite{gendelman2014normal} as the model systems for more accurate treatments, 
where the dissociation is safely defined.

\subsubsection{Fluctuating hydrodynamic description of \\the pressure-induced recovery}
We investigate how the pressure fluctuation changes the observed transport coefficients here.
Our numerical simulations 
suggest that there would be a pressure-induced 
mechanism 
of the 
recovery of Fourier's law 
in one dimensional momentum-conserving systems.
This non-Arrhenius 
temperature dependence is inconsistent with the scenario of thermally activated dissociation, so we should consider the recovery with 
 some continuum theory.
Here we consider the mechanism 
with
the fluctuating hydrodynamics.

One dimensional full fluctuating hydrodynamic equations describe the slow motion of the conserved quantities
 (: mass, momentum and energy) and consist of the following equations of continuity,
 writing $(\rho,v,e)$ as mass density, velocity and energy density,
\begin{eqnarray}
\left\{
\begin{array}{lll}
\partial_t \rho+\partial_x(\rho v)=0\\
\partial_t (\rho v) +\partial_x(\rho v^2+\sigma) =0
,\hspace{3pt} \sigma=P-\mu\partial_xv -s
\\
\partial_t e+\partial_xJ =0
,\hspace{3pt}J = e v+v\sigma-\kappa \partial_x T-g
\hspace{22pt}.
\end{array}
\right.
\end{eqnarray}
Here $(P,T)$ express pressure and temperature and are the functions of mass density and internal energy density ($\rho,e-\rho v^2/2$).
($\mu,\kappa$) are bulk viscosity and heat conductivity.
$(s,g)$ are random stress and random heat current.
The currents of momentum and 
those of
energy include dissipation terms 
which consist of 
the deterministic part expressed as the linear irreversible process
and the stochastic part satisfying fluctuation dissipation relation (FDR)~\cite{kubo2012statistical}.
$(s,g)$ are described by the white noise
 as a consequence of the central limit theorem, and satisfy the following relations expressing the local FDR, using $\langle\rangle$ as the noise average,
\begin{eqnarray}
\left\{
\begin{array}{lll}
\langle s(x,t)s(x^\prime,t^\prime)\rangle&=&2\mu T\delta(x-x^\prime)\delta(t-t^\prime)
\\
\langle g(x,t)g(x^\prime,t^\prime)\rangle&=&2\kappa T^2\delta(x-x^\prime)\delta(t-t^\prime)
\\
\langle s(x,t)g(x^\prime,t^\prime)\rangle&=&0.
\end{array}
\right.
\end{eqnarray}
In addition,
the high wave number cutoff $\Lambda$ is assumed because of the non-continuum area of short-wavelength.

We 
want to derive some simple model to study the effect of compressibility based on these equations here. 
We treat our equilibrium result as the observation of the anomaly suppression, then 
do not consider the nonequilibrium  effects. 
Our treatment to discuss the convergence
also neglects the nonlinear terms of dissipation terms, 
because they are irrelevant even in the inviscid scaling~\cite{spohn2015fluctuating}.
We restrict our attention in such a parameter range.
We write $(\rho,u:=\rho v, e)$ as mass density, momentum density and energy density, 
and study the fluctuations$(\delta \rho,u,\delta e)$ from the base of a static equilibrium 
state $(\rho_0,0,e_0)$. 
Their motion is described by the following equations, 
at
the lowest order of nonlinearity neglecting irrelevant dissipative nonlinear terms, 
\begin{eqnarray}
\left\{
\begin{array}{lll}
\partial_t\delta \rho +\partial _x u = 0
\\
\partial_t u +\frac 1{\rho_0} \partial _x u^2
\\
 = 
-\left(\frac{\partial P}{\partial \rho}\right)_0 \partial _x\delta \rho
-\left(\frac{\partial P}{\partial \rho}\right)_0 \partial _x(\delta e-\frac {u^2}{2\rho_0})
+\nu_0\Delta u+\partial_x s_0
\\\hspace{10pt}
-\frac 1 2 \partial_x\left[
\left(\frac{\partial^2 P}{\partial \rho^2}\right)_0 (\delta \rho^2)
+2 \left(\frac{\partial^2 P}{\partial \rho\partial e}\right)_0 \delta \rho\delta e
+\left(\frac{\partial^2 P}{\partial e^2}\right)_0 \delta e^2
\right]
\\
\partial_t\delta e +\frac{1}{\rho_0}\partial _x (u\delta e)
\\ =
 -h_0\partial_x u 
+\kappa_0\left(\frac{\partial T}{\partial \rho}\right)_0 \Delta\delta \rho
+\kappa_0\left(\frac{\partial T}{\partial e}\right)_0 \Delta\delta e
+\partial_x g_0
\\
\hspace{10pt}
-\frac{e_0}{\rho_0}\partial_x\left(u \frac{\delta\rho}{\rho_0}\right)
-\partial_x
\left[
u/\rho_0
\left(
\left(\frac{\partial P}{\partial \rho}\right)_0 \delta \rho
+\left(\frac{\partial P}{\partial \rho}\right)_0 \delta e
\right)
\right] \,\,.
\end{array}
\right.
\nonumber\\
\end{eqnarray}
We defined ($\nu_0:=
\mu_0/\rho_0
, h_0:=(P_0+e_0)/\rho_0$)
and $0$ is the indicator of 
original equilibrium values. 
The derivation of the above equations is almost the same with
~\cite{spohn2015fluctuating}.
He chose the particle distance as a conserved quantity instead of mass, 
but this is basically equivalent~\cite{spohn2014nonlinear}.
The nonlinear terms arise from pressure nonlinearity 
and Galilean invariance of fluids. Pressure nonlinearity is well studied in his papers.
He assumed fluid equations in the particle currents (Lagrangian description) and no streaming terms appear in the decription. Compared with pressure, streaming terms seem to have attracted less attention in spite of the famous mechanism of the anomaly induced by them~\cite{forster1977large}. 
Then we continue the analysis with the approximation of weak pressure fluctuations here
to study the other origin of nonlinear effects. 
For further approximation, we treat the pressure fluctuations as the perturbations 
and neglect the corresponding nonlinear terms. 
We also neglect the nonlinear term including ($\delta\rho/\rho_0$)
taking into account the arbitrarily large lattice constant of lattice sytems, 
 then get the following equations,
\begin{eqnarray}
\left\{
\begin{array}{lll}
\partial_t\delta \rho+\partial_x u=0
\\
\partial_tu +\frac 1 {\rho_0}\partial_x u^2
=-Y_0\partial_x \delta \rho -Z_0\partial_x \delta e
+\nu_0\Delta u+\partial_x s_0
\\
\partial_t\delta e +\frac 1 {\rho_0}\partial_x u \delta e 
=-h_0 \partial _x u 
+ D_0\Delta \delta e + E_0\Delta \delta \rho +\partial_x g_0 \,\,.
\end{array}
\right.
\nonumber\\
\label{eq:original}
\end{eqnarray}
Here we defined the coefficients
$Y_0:=\left(\frac{\partial P}{\partial \rho}\right)_0$, 
$D_0:=\kappa_0\left(\frac{\partial T}{\partial e}\right)_0$,
$Z_0:=\left(\frac{\partial P}{\partial \rho}\right)_0$ and
$E_0:=\kappa_0\left(\frac{\partial T}{\partial \rho}\right)_0$.
$D_0$ is positive in general.
$h_0$ can take the negative values, e.g. negative $h_0$ corresponds to some expanded states in FPU-$\beta$.
In $Z_0=0$ case, corresponding to zero pressure in FPU-$\beta$, $Y_0$ is positive because the sound velocity takes the real value. 
Particularly, ($Y_0,Z_0,h_0,E_0\to 0$) corresponds to the dynamics
 of the passive scalar on noisy Burgers fluids~\cite{drossel2002passive}, and the divergence of 
observed transport coefficients ($\nu, D$) obeys the inviscid scaling~\cite{forster1977large,kardar1986dynamic}.
Then, one can see this model as a minimal model to 
study the effect of compression. 

The explicit calculation of renormalization into the observed transport coefficients 
on these equations clarifies how the pressure fluctuations affect the observed heat conductivity as a consequence of Galilean invariance. 
In particular, $Z=0$ ($Z$ negligible) case 
is
consistent with 
{\it decoupling hypothesis} in~\cite{spohn2015fluctuating}, where the heat mode is passive to the sound modes, 
and we choose the parameter.
Here, those waves are defined as the plane waves of perfect fluids. 
The propagation speed is zero in heat mode and the sound speed in sound modes. 
Original decoupling hypothesis assumes that relation between plane waves only at the long wavelength and means the asymptotic irrelevance of contribution from the heat to the sounds.
The prediction based on the approximation is tested with good correspondence~\cite{das2014numerical}, 
so this hypothesis would capture the truth. We also emphasize $Z=0$ is the unique parameter satisfying decoupling hypothesis except special sets of the coefficients.
We put the MCT results
under this parameter choice, 
which renormalizes the contribution of the higher wave length fluctuations at once in the lowest order of nonlinearity
~\cite{forster1977large}.
The result of heat diffusion coefficient takes the following value in the long wavelength and low frequency limit,
\begin{eqnarray}
\frac{D_R}{D_0}&=&1+\frac{T_0}{2\pi\rho_0\sqrt{Y_0}\sqrt{D_0(D_0+\nu_0)}}
\frac{D_0+2\nu_0}{D_0+\nu_0}
\nonumber\\
&&\times\left(
2\arctan(W_0\Lambda)
+\frac{(3D_0+\nu_0)W_0\Lambda}{(D_0+2\nu_0)(1+(W_0\Lambda)^2)}
\right),
\nonumber
\\W_0&:=&\sqrt{\frac{D_0(D_0+\nu_0)}{Y_0}}.
\label{eq:obsdr}
\end{eqnarray}
This convergence is consistent with our numerical simulation.
We note 
that the zero sound velocity limit ($Y_0\to0$) recovers the ordinary result of the $D_R$ divergence ($D_R\to\infty)$.
Considering the difference between ours and noisy Burgers solution ($Y_0\to 0$ case), one can understand the convergence of the renormalized heat conductivity 
($\kappa_R\simeq D_R/\left(\frac{\partial T}{\partial e}\right)_0$) as the 
cutoff of the anomaly around the characteristic wavenumber $k_*:= 2\sqrt{Y_0}/\nu_0$.
This wave number is defined as the characteristic wavenumber of the momentum field 
Green function absorbing the density fluctuations, given by
\begin{eqnarray}
G_0(k,t):=\int \frac{d\omega e^{ikt+i\omega t}}{i\omega+\nu_0k^2-iY_0k^2/\omega},
\end{eqnarray}
to show the effect of compressibility, i.e.
\begin{eqnarray}
G_0(k,t)\simeq
\left\{
\begin{array}{ll}
e^{-\nu_0k^2 t} ,&k\gg k_*
\\
e^{-\frac {\nu_0k^2} 2 t}\cos(\frac{\nu_0k_*|k|} 2 t),& k\ll k_* .
\end{array}
\right.
\end{eqnarray}
Then the compressibility would be intrinsic for this convergence. 
As other results, we derived the followings,
\begin{eqnarray}
\frac{\nu_R}{\nu_0}&=&\infty,
\\
\frac{h_R}{h_0}&=&1+
\frac{(D_0-E_0/h_0)T_0\arctan(W_0\Lambda)}{2\pi\nu_0\rho_0\sqrt{Y_0}\sqrt{D_0(D_0+\nu_0)}}
,
\end{eqnarray}
and $Y_R-Y_0,Z_R-Z_0,E_R-E_0=0$.
 $h_R\neq h_0$ is understood as the renormalization of 
the thermal fluctuations into the energy.
The value of the renormalized viscosity $\nu_R\to\infty$ asserts the result of $D_R$ is caused by some 
breakdown of the hyperscaling
because two transport coefficients show the same scaling with sizes if the hyperscaling holds~\cite{forster1977large}.

However, there are at least three problems
for this interpretation. 
One is the unknown parameter $\nu_0$. 
The above result of $D_R$ asserts that the cutoff would be determined by the balance between viscosity and compressibility, but $\nu_0$ is not observable.
The second one is the correspondence of this result and the previous works 
which claim the divergence of heat conductivity in the inviscid limit.
The third one is $\nu_R$ divergence.
Some RG flow study can avoid the problems,
but the RG study of those equations lacks brevity in spite of 
the unclear validity of our approximation.

The passive scalar limit, $Z_0=h_0=E_0=0$ 
extremely simplifies the problem, then we study the RG flow on this condition 
with believing some universality.
This parameter simplification does not change our MCT results for the 
two transport coefficients $D$ and $\nu$. It means the results of our MCT analysis can be understood based on the following RG discussion.
For further simplicity, we do not discuss the slightly messy flow of passive scalar here.
Its flow does not affect the flow of the other conserved fields and should 
show the breakdown of the hyperscaling as the consistency with our MCT analysis (i.e. the value is absorbed into 0 under the flow).
Then our starting point is
\begin{eqnarray}
\left\{
\begin{array}{lll}
\partial_t\delta \rho+\partial_x u&=&0
\\
\partial_tu +\frac 1 {\rho_0}\partial_x u^2&=&-Y_0\partial_x \delta \rho
+\nu_0\Delta u+\partial_x s_0
\end{array}
\right..
\end{eqnarray}
These equations correspond to fluctuating hydrodynamic equations where pressure is 
temperature-independent and weakly fluctuates. $c_0:=\sqrt{Y_0}$ is the original sound velocity.
One can do the same diagram calculation with the noisy Burgers case~\cite{forster1977large}, then we briefly report the results.
We introduce the formal nonlinear intensity $\lambda$ and replace $1/\rho_0$ with $\lambda_0:=1/\rho_0$.
We write the noise intensity as $\Sigma,\,\Sigma_0 := 2\rho_0\nu_0 T_0$ to avoid the confusion of notation 
and define the coefficients $\bar Y:=4Y/(\nu\Lambda)^2,\bar\lambda:=
\lambda\sqrt {\Sigma/(\nu^3\Lambda)}
$ for preparation. 

\begin{figure}[tbp]
  \includegraphics[width=80mm]{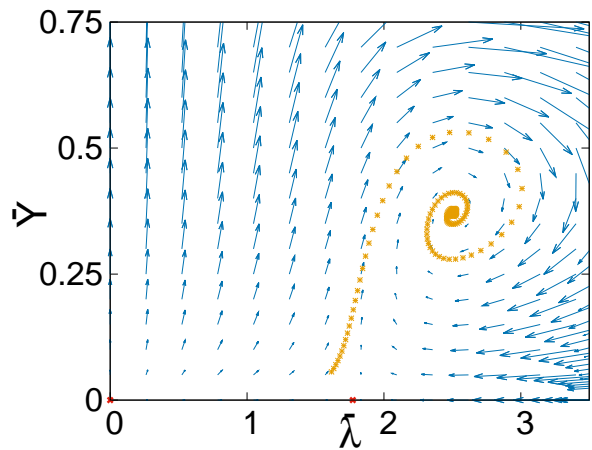}
 \caption{
The RG flow field (\ref{eq:RG1}),(\ref{eq:RG2}) of approximated full fluctuating hydrodynamic equations in the passive scalar limit. 
There are four fixed points, $(\bar\lambda,\bar Y)=(0,0),(0,\infty),(\sqrt\pi,0),
(\sqrt{2\pi},1-2^{-2/3})$.
The latter two points are the well-known inviscid fixed point and 
the non-trivial ballistic fixed point.
The linearly stable fixed point is the ballisitc fixed point only.
The inviscid fixed point is unstable except in $\bar Y=0$.
We marked the three fixed points taking finite values and show the typical flow in the same figure.
$\bar Y$ slowly increases around the inviscid fixed point and 
the flow is accelerated after the detachment and wraps around the ballistic fixd point.
 }
 \label{fig:five}
\end{figure}

The renormalization of the contribution from short wavelength 
$\Lambda>|k|>\Lambda e^{-l}, u^>:=
u\theta(|k|-\Lambda e^{-l})$ to 
long wavelength $|k|<e^{-l}, u^<:=
u-u^>$ and the rescaling along the scaling of $u$,
$k^\prime:=e^lk,\omega^\prime:=$
$e^{\int^l_0 z(l^\prime)dl^\prime}$
$\omega,u^<=:\zeta u^\prime$,$y(l):=\frac{d}{dl}\log\zeta-(z+1/2)$
shape the following RG flow,
\begin{eqnarray}
\frac{d\nu}{dl}&=&\nu\left(z-2+\frac{\bar\lambda^2}{2\pi}\right)
\\
\frac{d\Sigma}{dl}&=&\Sigma\left(z-2-2y+\frac{\bar\lambda^2}{2\pi}
\frac 1 {\sqrt{1-\bar Y}^3}
\right)
\\
\frac{d\lambda}{dl}&=&\lambda(-3/2+z+y)
\\
\frac{dY}{dl}&=&Y(-2+2z).
\end{eqnarray}
$\zeta$ is chosen to make the same RG flow of $(\nu,\Sigma)$,
and $z$ is chosen as $\nu$ fixed at their initial values.
The values are
\begin{eqnarray}
y&=&\frac{\bar \lambda^2}{4\pi}\left[
\frac 1 {\sqrt{1-\bar Y}^3}-1
\right]
\\
z&=&2-\frac{\bar\lambda^2}{2\pi}
.
\end{eqnarray}
Under the choice, the RG flow is reduced to 
\begin{eqnarray}
\frac{d\bar Y}{dl}&=& \bar Y\left(
2-\frac{\bar\lambda^2}{\pi}
\right)
\label{eq:RG1}
\\
\frac{d\bar \lambda}{dl}&=& \bar\lambda
\left[
\frac 1 2 -\frac{\bar\lambda^2}{4\pi}
\left(
3- \frac 1 {\sqrt{1-\bar Y}^3}
\right)
\right].
\label{eq:RG2}
\end{eqnarray}
We put the flow in Figure \ref{fig:five}.
Now one can find four fixed points,
trivial Gaussian fixed point
$(\bar \lambda,\bar \lambda,y,z)=(0,0,0,2)$,
the well-known inviscid fixed point
$(\bar\lambda,\bar Y,y,z)=(\sqrt{\pi},0,0,3/2)$,
$\bar Y$ divergent Gaussian fixed point
$(\bar\lambda,\bar Y,y,z)=(0,\infty,0,2)$
and the other non-trivial fixed point
$(\bar\lambda,\bar Y,y,z)=(\sqrt{2\pi},1-2^{-2/3},1/2,1)$.
Here we call it {\it ballistic} fixed point because of its $z$ value $z=1$. 
Gaussian fixed points are unstable in one dimension as already pointed out~\cite{forster1977large}, and 
the situation is the same in this analysis.
It is noteworthy that only the ballistic fixed point is linearly stable.
$\bar Y$ divergent Gaussian fixed point is unstable in $\bar\lambda>0$, so it would have no meaning because we started the analysis with $\bar\lambda_0\neq 0$.
The inviscid fixed point is unstable to the $\bar Y$ direction and
stable only in $\bar Y=0$. 
If the system at the point is perturbed to the $\bar Y$ direction, $\bar Y$ increases and 
the flow is absorbed to the ballistic fixed point. This result corresponds to the anomaly cutoff 
in our MCT analysis.
We repeat the positivity of $\bar Y_0$ in this case.

With the above results, one can interpret the anomaly cutoff of the observed heat conductivity as some crossover between those two fixed points. 
The cutoff is determined by the characteristic wavenumber $\Lambda_*$ where 
viscosity and compressibility are balanced. $\Lambda_*$ is defined as the 
minimum wavenumber satisfying the following relation,
\begin{eqnarray}
\frac{2c}{\nu(\Lambda_*)\Lambda_*} = 1.
\end{eqnarray}
These quantities are observable. If the system obeys the inviscid scaling, 
one can roughly estimate $\nu$ as $\nu(k)\sim k^{-1/3}$ then the viscous fluids 
have such $\Lambda_*$ because of $c>0$. 
The corresponding system size $N_*:=2\pi/\Lambda_*$ is also given if $\Lambda_*$ exists. 
Now, the scenario asserted from the results is 
as follows.
At first, heat conductivity increases obeying the inviscid scaling in the small system sizes
$N\ll N_*$, but the trend changes at $N\sim N_*$ and the heat conductivity shows the convergence
at the larger system sizes $N\gg N_*$, i.e.
\begin{eqnarray}
\kappa\sim\left[\min[N,N_*]\right]^{1/3}.
\end{eqnarray}
The flow stagnates around the inviscid fixed point, then there can be a certain period to show the anomaly. The result is consistent with the result of c-FPU-$\beta$ at ($l=4,T\gg1$) 
where bulk heat conductivity increases with $\alpha\sim1/3$ transiently and shows the tendency of saturation at larger system sizes. 
The result at $l=2$ showing the rapid saturation can be interpreted as the case where 
initial $(\bar \lambda,\bar Y)$ are near the ballistic fixed point. 

The above scenario would be appropriately modified to include the apparent counter example, the inviscid fixed point $(\bar Y,\lambda)=(0,\sqrt{\pi})$ in our analysis.
This parameter $\bar Y=0$ corresponds to
the $l=0$ case showing the agreement to the
ordinary theories~\cite{narayan2002anomalous,spohn2015fluctuating}.
The stability of two nontrivial fixed points would be also changed 
by the effect of pressure nonlinearity. 
Even though, we expect full analysis would keep the main ideas of this discussion 
that the convergence our simulation captured would be connected to the crossover between two fixed points in the intermediate wavelength. 
According to our RG analysis, there can be a transition in the long wavelength limit of full fluctuating hydrodynamics.

\section{Discussions}
Here we discuss the correspondence between our results and previous works, the RG flow results
(explicit calculations of 
noisy incompressible fluids, those of noisy Burgers fluids~\cite{forster1977large}
 and the scaling analysis~\cite{narayan2002anomalous}), 
MCT results of Full FH (for the scaling of equilibrium correlators~\cite{spohn2015fluctuating} and
for steadily sheared systems~\cite{lutsko2002long}), 
numerical simulations and the carbon-nanotube experiments~\cite{chang2008breakdown}.

Thermally activated dissociation and pressure fluctuations would 
cause the recovery of Fourier's law
in one dimensional momentum-conserving systems.
The integrable systems at the dilute (high-energy) limit 
(e.g. PR-$\delta$~\cite{savin2014thermal}
 and soft rod~\cite{gendelman2014normal}) would be the models of the 
thermally activated recovery.
The poly(quasi)-stable interaction potential systems like coupled rotator model would be also in this class 
by the same reason. 
The direct calculation of heat conductivity in coupled rotator model with onsite potential predicts the Arrhenius $T$-dependence~\cite{pereira2006normal}.
With the same mechanism, carbon-nanotube may also show the saturation caused by thermally generated defects
which is not avoidable in the real experiments~\cite{kittel2005introduction}.
The dissociation discussed here is not static and did not get the sufficient test,
and the realtion between dissociation class and nearly integrable systems which show the convergence (e.g. diatomic hardcore with small mass difference~\cite{chen2014nonintegrability}) is still unclear.
Further study is needed. 
Also, we note one can see the long time tails of energy currents in diatomic PR-$6$ at the golden mass ratio. The reason would be the nonintegrability in its dilute limit (diatomic hardcore). 
At the same time, the sum of momentum in each same-mass particles shows the long time tail, which means this observable is effectively conserved.
The diatomic systems may need careful treatments~\cite{chen2014nonintegrability,hurtado2015violation}.
The previous numerical simulation reporting the importance of the ballistic behavior in normal heat conduction~\cite{li20151d} was criticized because of their 
zero-temperature simulations~\cite{kosevich2015confinement}, but our following ballistic scaling observation is at the high temperature, where the system shows the anomaly at zero pressure.
Potential asymmetry makes non-zero pressure, 
then this ballistic-inviscid crossover on the RG flow may give some insight to the controvertial 
systems of the convergence 
like FPU-$\alpha,\beta$ and LJ~\cite{lepri2015heat,chen2012breakdown,wang2013validity,savin2014thermal,das2014heat,lepri2005one}.
For the real experiments, if our discussions are right, the dominance of these two 
effects is not clear, then to find the system showing the two 
mechanisms (corresponding to two temperature dependence) is also of interest.
The multi-particle collision system~\cite{di2015anomalous} is intriguing.
The 
recovery of Fourier's law 
in the thermodynamic limit should also still be discussed.

The continuum suppression of the anomaly is 
caused
by the response to the pressure fluctuations in our analysis,
and then compressibility can be an intrinsic factor. 
Our analysis considers the pressure fluctuations 
as the base of the linearized hydrodynamics and the wave number of saturation 
is decided with the sound velocity ($\sqrt{Y_0}$).
In incompressible fluids,
the pressure effect is the first order of nonlinearity (then $\mathcal O(l)$)
and  the response to the pressure becomes the second order (then $\mathcal O(l^2)$), 
so the (first derivative) RG flow of incompressible fluids~\cite{forster1977large} does not include this saturation.
In zero sound velocity limit, our result corresponds to 
the noisy Burgers result~\cite{forster1977large,kardar1986dynamic}.
We repeat our RG result 
demonstrated 
the instability of the inviscid fixed point to the assignment of compressibility. 
The property and the validity of our equations in $d>2$-dimensional cases, where 
the RG flow of Noisy Burgers has an unstable fixed 
point~\cite{forster1977large}, should be tested. 
It may be constructive to derive 
the ballistic fixed point along the discussion
deciding the anomalous exponent~\cite{narayan2002anomalous} here. 
Firstly, we rescale time and space as $x=x^\prime e^l, t=t^\prime e^{zl}$ 
and assume the conserved fields we choose $(\rho,{\bf u},e)$ take the same scaling
$(\rho,{\bf u},e)=\zeta (\rho^\prime,{\bf u}^\prime,e^\prime)$. 
Then, if mass conservation law
 $\partial_t \rho+{\bf \nabla  u}=0$ is required to be invariant under the RG flow,
$t$ and $x$ scale in the same way because of the required assumption. This is the desired exponent $z=1$. This scaling discussion is in contrast to the discussion in~\cite{narayan2002anomalous} 
which required the same rescaling of the set $(\rho,{\bf v},e)$, where the scaling ($z=3/2,d=1$) is derived from the requirement of the same scaling of  $(\rho,{\bf v},e)$, relevance of nonlinearity in the flow and the scaling invariance of the equal-time equilibrium fluctuations 
$\int dx^d(\rho^2,{\bf v}^2,e^2)$
under the flow (long range Gaussian property in their terms).
In their discussions, the scaling dimension of the three variables is $-d/2$ and 
it corresponds to $-(d/2+z)$ if the variables are treated as densities of the time directions (as a roughly sketched example:$\rho\to\rho\Delta t$) as described in \cite{forster1977large}.
The second assumption is satisfied in $d<2$ dimension.
The last assumption can be understood as the extensive property of the fluctuations. 
Then, in other words, one can get the ballistic fixed point
if one replaces the extensive fluctuations of $(\rho,{\bf v},e)$ with those of $(\rho,{\bf u},e)$ in their discussion. 
We also note that the scaling exponent $z =1$ is {\it independent of the dimension} in this discussion. 
We do not know the meaning.
Our scaling discussion expanded here is of heuristics and apparently has no use for the stability, but our example of the RG flow showed the possibility that compressibility should make the flow around the inviscid fixed point unstable.

In addition, the MCT analysis under sheared conditions (nonequilibrium steady states) already reported the tail strengthening, corresponding to the normal transport~\cite{lutsko2002long}.
This may work as an analogy of the possibility that the nonequilibrium effect causes the convergence.
They reported that the tail exponent changed at the characteristic wave number of the shear
and that the strengthened exponent of the long wave length corresponded to the normal transport. 
We did not discuss such possibility here, 
but it is not hard to suppose that the nonequilibrium condition (isobaric but varying temperature) may affect the heat conduction because of the global coupling as in the GK formula~\cite{casati2003anomalous}.

At the last, we go back to the most detailed MCT analysis of equilibrium correlations in 
one dimensional full FH~\cite{spohn2015fluctuating} for the discussion of the 
correspondence. 
We mention the difference that our analysis and his analysis are based on different equations. 
He started the analysis from the equations of continuity of conserved quantities
(volume, momentum and total energy), 
study the effect of pressure nonlinearity with MCT and with some approximations, 
and get the scalings of the autocorrelation functions for the
three 
plane
waves of the linearized perfect fluids (one heat mode and two sound modes). 
Their analysis is based on the linearized hydrodynamics added pressure nonlinearity in.
In his analysis, the anomaly is directly connected with the heat-heat mode correlator 
 which becomes to the symmetric Levy distribution in sound cones propagating 
at the sound speed. 
His prediction is already tested in some anomalous systems with good agreements, but some deviation 
was also reported~\cite{das2014numerical}.
He used some approximations, however, he also claimed 
the possibility that his result remains without his approximations in his equations.
His claim looks based on the mirror symmetry of sound cones and nonlinear terms of currents. 
Then the difference with us may come from the starting point equations.
Their equations do not include no streaming terms because of the derivative along the constituent particles~\cite{spohn2014nonlinear}, then the nonlinear currents are 
mirror-symmetric because the pressure depends only on mass and on internal energy. However, our equations include streaming terms and the nonlinear current ($u\delta e$) is mirror-antisymmetric. 
Our result may assert the significance of this difference.
One can have some questions to the validity of our $D_R$ convergence, so to get the credit, we also note the result of $D_R$ (\ref{eq:obsdr}) is also the exact solution of 
the passive scalar diffusion coefficient in the fluids neglecting their temperature dependence of pressure. As another derivation, one can easily get the same diagram of (\ref{eq:original}) with the same procedure of~\cite{bedeaux1974renormalization}.
Then our result has the universality based on the ballistic fixed point in some extent. 
 
To clarify the difference, the test of current fluctuations would be sufficient. 
His concrete prediction is as follows.
Firstly, he decomposed the conserved quantities
${\bf Q}_n=^T(\Delta x_n:=x_n-x_{n-1},p_n,\tilde e_n:=V(\Delta x_n)+p_n^2/(2m))$ into the three plane waves of the linearized perfect fluids $\tilde{\bf Q}_n$.
If one obtains the linearized equations of perfect fluids $(\partial_t+\partial_xA){\bf Q}=0$ and the covariance matrix $C \,s.t.\,C_{mn}:=\langle Q_m(x,t):Q_n(y,s)\rangle^{eq}_{T,P}/(\delta(x-y)\delta(t-s))$, 
one can define the linear transformation matrix $R$ satisfying the conditions $RAR^{-1}=diag (-c,0,c),RCR^T = diag(1,1,1)$
and derive $\tilde{\bf Q} :=R{\bf Q}$.
We used $\langle:\rangle^{eq}_{T,P}$ as $\langle f:g\rangle^{eq}_{T,P}:=\langle fg\rangle^{eq}_{T,P}-\langle f\rangle^{eq}_{T,P}\langle g\rangle^{eq}_{T,P}$.
One can decide $R$ uniquely except the trivial arbitrariness $\pm 1$.
One of the 
sets 
takes the following matrix.
See~\cite{spohn2014nonlinear} for details.
Here we abbreviate the time and the position for discussing the correlations of equal time and equal place. 
Shortening $\langle\rangle^{eq}_{T,P}$ to $\langle\rangle$, $\Delta x_n$ to $x$ and $V(\Delta x_n)$ to $V$,
and defining 
\begin{eqnarray}
\tilde \Gamma^{-1}&:=&T^{-1}\left(\langle x:x\rangle\langle V:V\rangle
-\langle x:V\rangle^2\right)
\nonumber
\\&&+\frac T 2\langle x:x\rangle
\end{eqnarray}
and
\begin{eqnarray}
&&\left\{
\begin{array}{lll}
c&:=&\tilde\Gamma\left(\langle(V+Px):(V+Px)\rangle+\frac{T^2}2\right)
\\
\tilde Z_1&:=&\sqrt{2T}c
\\
\tilde\kappa&:=&\sqrt{2T\tilde\Gamma}
\\
\partial_lP&:=&-\tilde\Gamma\left(\langle V:(V+Px)\rangle+\frac{T^2}2\right)
\\
\partial_eP&:=&\tilde\Gamma\langle x:(V+Px)\rangle \hspace{3pt},
\end{array}
\right.
\end{eqnarray}
we can define $R$ as
\begin{eqnarray}
R&:=&\tilde Z_1^{-1}
\begin{pmatrix}
\partial_l P& -c & \partial_e P \\
\tilde \kappa P & 0 &\tilde  \kappa\\
\partial_l P& c & \partial_e P
\end{pmatrix}
.
\end{eqnarray}
These values are easily calculated with the numerical integration of the corresponding canonical values. 
And then he derived the analytic formulae of $\tilde{\bf Q}$ autocorrelators.
One of the heat mode becomes symmetric Levy, and ones of the sound modes become 
the sound speed propagating KPZ. 

\begin{figure*}[tbp]
   \includegraphics[width=160mm]{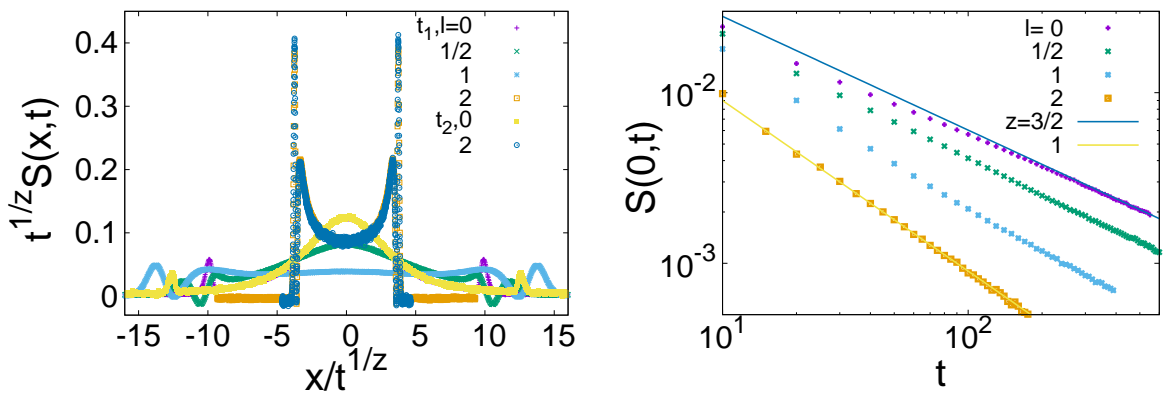}
  \caption{
Left: The equilibrium autocorrelation of the heat mode in c-FPU-$\beta$ at $T=1,t=t_n\sim400n/c, N=2048$.
Right: $S(0,t)$ in $t\leq t_2$.
z values of the left panel are chosen as 
$z=3/2$ 
at $l\neq 2$ and $z=1$ at $l=2$.
One can see the symmetric Levy in sound cones at $l=0$ and
 the growing bumps around the cones at $l\neq0$.
$S(x,t)$ shows the ballistic scaling at $l=2$.
The width scaling is the inviscid one $x/t^{2/3}$ at $l=0$ and the ballistic one $x/t$ at $l=2$.
$S(0,t)$ shows the clear crossover. 
  }
  \label{fig:six}
\end{figure*}

Here we report the deviation with $l$-increasing in c-FPU-$\beta$, which 
looks the asymptotic form  with respect to $t$-increasing. 
We set the data of the heat-heat correlator ($\Delta t =0.005,T=1$) in Figure \ref{fig:six}.
We abbreviate hear-mode autocorrelator $S_h$ as $S$ here.
 We only note about the sound modes that
inclined sound-sound correlators already reported in~\cite{spohn2015fluctuating} is also observed.
The symmetric Levy distribution of the heat-heat correlator in the sound cones $S(x,t)\simeq t^{-2/3}f_I(x/t^{2/3})$
are observed at $l=0$~\cite{das2014numerical},
however, the phase difference of the sound cones looks tripled at $l=0.5$.
The deviation is consistent with our other results and also with the previous reports~\cite{das2014numerical}. 
The height scaling looks to change at the same time. 
The observations of $S(0,t)$ show the change of the height scaling from 
$z=3/2$ to 
$z=1$. 
Furthermore, the heat-mode autocorrelator grows up around cones with the decreasing center top at $l=1,2$. The top is almost negligible at $l=2$.
We also note the bumps around the cones are scaled by $x/t$, which is just the ballistic scaling 
$z=1$.
This exponent change and growing bumps mean the scaling crossover from the inviscid scaling $S(x,t)\simeq t^{-2/3}f_I(x/t^{2/3})$ to the ballistic scaling $S(x,t)\simeq t^{-1}f_B(x/t)$.
The sound-heat correlations or the correlation overlaps~\cite{das2014numerical} may affect the result. If the anomaly remains in his equations, this difference may come from streaming terms. 
We could not observe the cutoff of  the scaling to provide the convergence, 
although the crossover is observed. 
It is difficult to observe the 
anomaly cutoff 
like 
$S(x,t)\simeq e^{-t/\tau}t^{-1}f_B(x/t)$ at the long time
scale 
as already pointed out in \cite{das2014role}
and we do not know how the possible cutoff looks to be.
We may need the extensive tests of his conjecture.
\section{Conclusion}
Our numerical experiments of steady heat conduction 
suggested some possible origins of normal heat conduction
in one dimensional momentum-conserving systems.
The recovery had been explained by the thermally activated dissociation.
However, compressed FPU-$\beta$ under the strong compression and of high energy ($l=2,4,T\gg1$)
 showed the recovery of 
Fourier's law
with the non-Arrhenius $T$-dependence of heat conductivity. 
It suggested the existence of the class of normal heat conduction the origin of which is explained by 
some macroscopic continuum theory in one dimensional momentum-conserving systems.
There would be at least two mechanisms of normal heat conduction.
The vanishing long time tails of 
total energy currents 
suggested some relation between 
the novel non-Arrhenius type 
recovery and the pressure. 
Then we tried to explain the two 
mechanisms with corresponding phenomenologies.
Firstly, for the 
case
of the Arrhenius $T$-dependence, 
we studied the coefficient with the transition-state theory following the scenario reported 
in~\cite{gendelman2014normal} that the 
convergence 
is caused by the thermal activation of dissociation and got quantitative agreement with our numerical simulations of PR-$6$ in a certain extent. 
Secondly, for the 
case of
the non-Arrhenius 
$T$-dependence, 
we 
executed 
the MCT analysis based on full FH equations of a specific 
limit 
and got the suggestion that the pressure fluctuations suppressed the anomaly. This was consistent with our numerical result of compressed FPU-$\beta$. 
The RG flow of further approximated equations asserted the connection between this result and the ballistic fixed point, another nontrivial fixed point from the inviscid fixed point. 
The fixed point was induced by the compressibility in our RG analysis.
Discussing the correspondence with the previous accurate theory of equilibrium correlators~\cite{spohn2015fluctuating}, we found the possibility of the inviscid-ballistic crossover of the heat mode autocorrelator.  
Further investigation is needed to examine the possibility of normal heat conduction in one dimensional momentum-conserving systems.
\begin{acknowledgements}
The author gratefully acknowledges helpful discussions with H. Hayakawa, K. Saitoh, S. Lepri and S. Takesue, and with S. Sasa and T. Hatano with the greatest thanks.
The discussions in International Seminar 2015 (YKIS2015): New Frontiers in Non-equilibrium Statistical Physics 2015 were the valuable opportunities for us.
\end{acknowledgements}

\appendix
\section{Derivation of the Green Kubo formula}
We derive the Green Kubo formula of the normal heat conduction here in two ways.
One of them (\ref{eq:GEGK})
 is the expansion from the globally equilibrium conditions 
and the other (\ref{eq:LEGK})
 is  the expansion from the locally equilibrium conditions.
In the far from equilibrium conditions, one cannot use the former, but can use the latter.

\subsection{Setting}
Let us consider a system that connects to two stochastic reservoirs at the edges.
Each reservoir has its connecting area and satisfies the local detailed balance condition.
We assume the time-reversal symmetry of the system Hamiltonian and the nearest-neighbor interaction (``short range'' interaction).
We further assume the unique steady state ensemble $P^{st}$ independent of the initial conditions.
Here we consider the macroscopic variables along the particle index, but can translate it to that along the field coordinate
$(A_i\to A(x)=\sum_i A_i\delta(x-x_i))$. Under that translation, the choice of the variable $r_i(:=x_{i+1}-x_i)$ in the expansion from the locally equilibriums would be translated into that of the mass density $\rho(x)=\sum_i m_i\delta(x-x_i)$. Also, one can extend this discussion to the case having additional order parameters.

We describe the fundamental relation to deive the formula as the preparation. We note $\Gamma$ as the phase space coordinate and $\mathcal H$ as the system Hamiltonian.
$h_i$ is the $i$-th particle energy density, $j_{i+1,i}$ is the corresponding energy current and $r_i$ is the $i$-th particle compression here.
$\beta_{\pm}(\beta_+>\beta_-)$ is the temperature of the reservoir $\pm$. 
Energy conservation law can be written as
\begin{eqnarray}
\partial_t h_i+j_{i+1,i}-j_{i,i-1} &=& \sum_{\pm}\dot Q ^\pm_i \chi[i\in D_\pm],
\end{eqnarray}
\begin{eqnarray}
\chi[\mbox{A}] &=& 
\left\{
\begin{array}{ll}
1& \mbox{A:true}\\
0& \mbox{A:false}
\end{array}.
\right.
\end{eqnarray}
$D_\pm$ is the connecting area of the reservoir $\pm$ ($|D_\pm|\ll N, \,N$:system size)
 and
$\dot Q^\pm_i$ is the energy came from the reservoir $\pm$ to the $i$-th particle per time.
By the local detailed balance condition, the path probability on the path $\hat \Gamma$ within the time $t\in[0,\tau]$ and that on the time-reversal path $\hat \Gamma^\dag$ 
satisfy the relation,
\begin{eqnarray}
\frac{P(\hat\Gamma|\Gamma_0)}{P(\hat\Gamma^\dag|\Gamma_\tau^*)}
= e^{-\beta_\pm Q_\tau^\pm}.
\label{eq:LDB}
\end{eqnarray}
Here, $Q_\tau^\pm$ is defined as $Q_\tau^\pm:=\int^\tau_0\sum_{i\in D_\pm} \dot Q^\pm_i$,
$\Gamma_0$ is the initial state of $\hat\Gamma$ and $\Gamma_\tau$ is the end state of $\hat\Gamma$.
$*$ is the index that expresses the time-reversal quantity $B^*$ of the instantaneous variable $B(\Gamma)$.
$P(\hat \Gamma|\Gamma)$ is the conditional probability of the path under this thermostat condition. 

It is convenient for later discussions to introduce McLennan ensembles (\ref{eq:McLE})~\cite{maes2010rigorous,itami2015nonequilibrium}.
Firstly, we note that 
fluctuation theorem (\ref{eq:FT}) holds for path-dependent variables $\mathcal A(\hat\Gamma)$
because of the existence of the local detailed balance(\ref{eq:LDB}).
Now we use $\langle \rangle $ as the average of the stochastic reservoir forces.
Defining 
\begin{eqnarray}
\Sigma(\hat\Gamma):=\beta_\pm Q_\tau^\pm+\log\frac{P^{ref}(\Gamma_0)}{P^{ref}(\Gamma_\tau^*)}
\end{eqnarray}
and using $\dag$ as the indicator of the time-reversals of the path-dependent variables ($\mathcal A^\dag(\hat\Gamma)=\mathcal A(\hat \Gamma^\dag)$),
one can get 
\begin{eqnarray}
\langle \mathcal A(\hat \Gamma)\rangle 
&:=&\int d\hat \Gamma P(\hat\Gamma|\Gamma_0)P^{ref}(\Gamma_0) \mathcal A(\hat \Gamma)
\\&=&
\int d\hat \Gamma^\dag P(\hat\Gamma^\dag|\Gamma_\tau^*)P^{ref}(\Gamma_\tau^*)
 \mathcal A(\hat \Gamma)
\nonumber\\&&
\hspace{10pt}\times \exp\left(\beta_\pm Q_\tau^\pm+\log\frac{P^{ref}(\Gamma_0)}{P^{ref}(\Gamma_\tau^*)}\right)
\\&=&
\int d\hat \Gamma P(\hat\Gamma|\Gamma_0)P^{ref}(\Gamma_0) e^{\Sigma^\dag(\hat \Gamma)}\mathcal A^\dag(\hat \Gamma).
\end{eqnarray}
That is 
\begin{eqnarray}
\langle \mathcal A \rangle 
=
\langle \mathcal A^\dag e^{\Sigma^\dag}\rangle. \label{eq:FT}
\end{eqnarray}
Secondly, we consider the time evolution of an ensemble $P_\tau$ that started from a reference ensemble $P^{ref}$.
We choose $\mathcal A=\delta(\Gamma^\prime-\Gamma_\tau)$ in the above fluctuation theorem, and get
\begin{eqnarray}
P_\tau(\Gamma^\prime)&=&
\langle \delta(\Gamma^\prime-\Gamma_\tau) \rangle 
=
\langle \delta(\Gamma^\prime-\Gamma_0^*) e^{\Sigma^\dag}\rangle
\\ &=&
P^{ref}((\Gamma^\prime)^*)
\nonumber \\ &&
\times
\int d\hat \Gamma P(\hat\Gamma|\Gamma_0)e^{\Sigma^\dag(\hat \Gamma)}
\delta(\Gamma_0-(\Gamma^\prime)^*).
\end{eqnarray}
We take the limit $\tau\to\infty$, and the steady distribution is expressed as
\begin{eqnarray}
P^{st}(\Gamma)&=&
P^{ref}(\Gamma^*)
\nonumber\\
&&\times
\lim_{\tau\to\infty} 
\int d\hat \Gamma P(\hat\Gamma|\Gamma_0)e^{\Sigma^\dag(\hat \Gamma)}
\delta(\Gamma_0-\Gamma^*).
\hspace{15pt}
\end{eqnarray}
We promise that $\langle \rangle^{st}$ expresses the steady state average.
This expression is transformed into another form for the instantaneous variables $B(\Gamma)$. Using the relation $d\Gamma=d\Gamma^*$, one can get the tractable form
\begin{eqnarray}
\langle B \rangle^{st}
&=&\int d\Gamma^* B^*(\Gamma^*)(P^{st})^*(\Gamma^*)
\\&=&\int d\Gamma B^*(\Gamma)(P^{st})^*(\Gamma)
\\&=&
\lim_{\tau\to\infty} 
\int d\hat\Gamma B^*(\Gamma)P^{ref}(\Gamma) P(\hat \Gamma|\Gamma)e^{\Sigma^\dag(\hat\Gamma)}.
\hspace{8pt}
\end{eqnarray}
Namely,
\begin{eqnarray}
\langle B \rangle^{st}
&=&
\lim_{\tau\to\infty} 
\langle B^*e^{\Sigma^\dag} \rangle^{ref}. \label{eq:McLE}
\end{eqnarray}
We expressed the reference ensemble average as $\langle\rangle^{ref}$.

\subsection{Derivation}
\subsubsection{Derivation-1. \\Expansions from globally equilibrium conditions}
We consider the expansion from the static globally equilibrium state 
\begin{eqnarray}
(\langle r_i\rangle^{st},\langle p_i\rangle^{st},\langle h_i\rangle^{st} ) =(l,0,e).
\end{eqnarray}
Corresponding intensive variables are 
\begin{eqnarray}
(\beta P,\beta V,\beta)=(\bar\beta P_0,0,\beta), \,\,\bar \beta =\frac{\beta_++\beta_-}{2}.
\end{eqnarray} 
$P$ is pressure and $V$ is averaged velocity.
Now we choose
\begin{eqnarray}
P^{ref}=P^{can}_{\bar\beta}:=e^{-\bar\beta\mathcal H}/Z
\end{eqnarray} 
as the reference ensemble.
It means the choice of $(T,l)$.
We use $\langle \rangle^{can}_{\bar\beta}$ to express the average by the ensemble $P^{can}_{\bar\beta}$.
Entropy production $\Sigma$ takes the form
\begin{eqnarray}
\Sigma &=& -\beta_\pm Q^\pm_\tau+\bar\beta(\mathcal H(\Gamma^*_\tau)-\mathcal H (\Gamma))
=\Delta \beta Q.
\end{eqnarray} 
We used the time-reversal symmetry of the Hamiltonian $\mathcal H(\Gamma^*_\tau)=\mathcal H(\Gamma_\tau)$ 
and defined $\Delta \beta:=\beta_+-\beta_-, Q := (Q_--Q_+)/2$.
Corresponding time-reversals are 
$\Sigma^\dag=-\Delta\beta Q, \dot Q^*=-\dot Q$.
The energy conservation law of the systems at the steady states requires $\sum_{\pm}\langle \dot Q^\pm\rangle^{st}=0$.
Also, the energy balance at the bulk ($n\bar\in D_\pm$) yields the relation, 
\begin{eqnarray}
|\langle J\rangle^{st}| &:=& |\langle j_{n+1,n}\rangle ^{st}| 
= \langle \dot Q\rangle^{st}
\\&=&
\lim_{\tau\to \infty} \langle -\dot Q e^{-\Delta\beta Q}\rangle^{can}_{\bar\beta}
\\ &=&
\Delta \beta \int ^\infty _0 dt\langle \dot Q (0)\dot Q (t)\rangle^{can}_{\bar\beta} +\mathcal O((\Delta\beta)^2).
\hspace{12pt}
\label{eq:GEGK}
\end{eqnarray}  
The sign of $J$ is minus if the reservoir $+$ is at the left and plus if it is at the right.
The error from the relation can be large if $\Delta \beta$ is comparable with $\bar\beta$ and not negligible in that case.
\subsubsection{Derivation-2. \\Expansions from locally equilibrium conditions} 
One can know the steady state value of the energy and the compression through the careful observation
even if their values spatially change,
and we consider the expansion from the static locally equilibrium conditions
\begin{eqnarray}
(\langle r_i\rangle^{st},\langle p_i\rangle^{st},\langle h_i\rangle^{st} ) =(l_i,0,e_i).
\end{eqnarray}
Spatial homogeneity of pressure is needed to keep the zero value of the momenta, then 
corresponding intensive variables are 
\begin{eqnarray}
((\beta P)_i,(\beta V)_i,\beta_i)=(\beta_i P,0,\beta_i).
\end{eqnarray} 
Now we suppose $l_i$ and $e_i$ are slowly varying quantities in space scaled with the correlation length of microscopic variables
(particularly $j_{i+1,i},v_i:=p_i/m_i$).
Also, we suppose the temperature gaps at the edges are negligible.
If the system shows the normal heat conduction, one can realize them with sufficiently large (but finite) system sizes $N$.
Under the condition, one can expand the contribution from the entropy production $\Sigma$ around the local equilibriums.
So we take the isobaric local equilibrium ensemble as the reference ensemble,
\begin{eqnarray}
P^{ref}=P^{lc}_{\{\beta_i\}_i,P}:=e^{-\beta_ih_i-\frac{\beta_{i+1}+\beta_i}{2}Pr_i}/Z. \label{eq:lce}
\end{eqnarray} 
It means the choice of $(\{T_i\}_i,P)$.
$\beta_i$ is chosen as the parameter satisfying the relation,
\begin{eqnarray}
i\bar\in D_\pm, \langle h_i\rangle^{lc}_{\{\beta_i\}_i,P}= \langle h_i\rangle^{st}
; \,\, i\in D_\pm ,\beta_i=\beta_\pm.
\end{eqnarray} 
We use $\langle\rangle^{lc}_{(\{\beta_i\}_i,P)}$ to express the average by the ensemble $P^{lc}_{\{\beta_i\}_i,P}$.
Entropy production takes the value 
\begin{eqnarray}
\Sigma &=& -\beta_\pm Q^\pm_\tau+\beta_i(h_i(\Gamma^*_\tau)-h_i (\Gamma))
\nonumber\\
&&\hspace{38pt}
+\frac{\beta_{i+1}+\beta_i}2P(r_i(\Gamma^*_\tau)-r_i (\Gamma))
\\ &=&
-\int ^\tau_0dt \left[\beta_i(j_{i+1,i}-j_{i,i-1})
\right.\nonumber\\&&\hspace{38pt}\left.
-\frac{\beta_{i+1}+\beta_i}2P(v_{i+1}-v_i)\right]
\\ &=&
\int ^\tau_0dt \left[(\beta_{i+1}-\beta_i)j_{i+1,i}-\frac{\beta_{i+1}-\beta_{i-1}}2Pv_i\right].\hspace{18pt}
\end{eqnarray} 
The time-reversals are $\Sigma ^\dag=-\Sigma,j^*_{i+1,i}=-j_{i+1,i}$.
Then the averaged value of the current at the bulk $n\sim N/2$ can be estimated as
\begin{eqnarray}
&&\langle J \rangle^{st}:=
\langle j_{n+1,n}\rangle^{st}
\\&=&
\lim_{\tau\to \infty} \left\langle - j_{n+1,n}
\right.\nonumber\\&&\hspace{20pt}\left.\times e^{-\int ^\tau_0dt [(\beta_{i+1}-\beta_i)j_{i+1,i}-\frac{\beta_{i+1}-\beta_{i-1}}2Pv_i]}\right\rangle^{lc}_{(\{\beta_i\}_i,P)}
\\ &=&
\left\langle  j_{n+1,n}(0)
\int ^\infty_0dt \left[(\beta_{i+1}-\beta_i)j_{i+1,i}
\right.\right.\nonumber\\&&\hspace{10pt}\left.\left.
-\frac{\beta_{i+1}-\beta_{i-1}}2Pv_i\right]\right\rangle^{lc}_{(\{\beta_i\}_i,P)}
+\mathcal O ((d\partial_n\beta)^2).
\end{eqnarray}
$d$ is the largest correlation length of the currents. 
$\langle j_{n+1,n}\rangle^{lc}_{(\{\beta_i\}_i,P)}=\langle j^*_{n+1,n}\rangle^{lc}_{(\{\beta_i\}_i,P)}=0$ follows the time reversal symmetry of the chosen ensemble reflecting $\mathcal H^*=\mathcal H$.
This relation corresponds to that derived by Casati \& Prosen~\cite{casati2003anomalous} under the chaotic pictures.
However, the estimated error under the anomalous heat conduction is different from the above. Under the condition, the estimated error was $\mathcal O ((N\partial_n\beta)^2)=\mathcal O ((\Delta\beta)^2)$ because there were no scale separations anymore and the correlation length grows to the system size ($d\sim N$).
The situation of the system of normal heat conduction is in sharp contrast to it.
In the systems of normal heat conduction, this modified error estimation 
$\mathcal O ((d\partial_n\beta)^2)$
assures us the safety to use the relation even under the far from equilibrium conditions ($\Delta \beta \sim\bar \beta$).
Our interest is in the Green Kubo formula on the systems of normal heat conduction, so the average of the right hand side can be estimated under the equilibrium condition with further expansions.
Writing the ensemble average by the isobaric canonical $P^{can}_{\beta,P}:=\exp(-\beta (\mathcal H+P\sum_ir_i))/Z$ as $\langle\rangle^{can}_{(\beta,P)}$, we get the estimation,
\begin{eqnarray}
\langle J \rangle^{st}&=&
(\beta_{n+1}-\beta_n)
\nonumber\\&&\times\left\langle  j_{n+1,n}(0)\int^\infty_0 dt \sum_i [j_{i+1,i}-Pv_i]\right\rangle^{can}_{(\beta_n,P)}
\nonumber\\&&+\mathcal O ((d\partial_n\beta)^2,d^2\partial^2_n\beta).
\end{eqnarray}
We promise that we do not take the summation on the index $n$ (though we have taken the summation on the index $i$).
This relation is valid for each gradient $\beta_{n+1,n} (n\sim N/2)$,
i.e. $\beta_n$-dependence of the averaged current correlations corresponds to the temperature dependence of the heat conductivity $(\times \beta^{-2})$ in the isobaric cases.
Also, at the middle of the system $n\sim N/2$, the forces from the reservoirs within the correlation time $\tau_c$ does not affect the dynamics of the particles far from the edges within the time
 and the contribution of the past forces from the reservoirs are used only for the setting of the initial condition, 
then one can estimate the average with static isolated-periodic boundary.
This translation is done by the change of the conditional probability $P(\hat\Gamma|\Gamma)\to\delta(\hat\Gamma(\Gamma)-\hat\Gamma)$
(, where $\hat\Gamma(\Gamma)$ is the solution of the equation of motion determined by the system Hamiltonian with the initial condition $\Gamma$,) and we note it as $\langle\rangle^{can-IP}_{(\beta,P)}$.
Then it can be reduced to
\begin{eqnarray}
\langle J \rangle^{st}&=&
(\beta_{n+1}-\beta_n)
\nonumber\\&&\times\left\langle  j_{n+1,n}(0)\int^\infty_0 dt \sum_i [j_{i+1,i}-Pv_i]\right\rangle^{can-IP}_{(\beta_n,P)}
\nonumber\\&&
+\mathcal O \left((d\partial_n\beta)^2,d^2\partial^2_n\beta,d\partial_n\beta \frac{c\tau_c}N\right).
\label{eq:LEGK-d}
\end{eqnarray}
We repeat the no summation on the index $n$.
The propagation speed of the effect is constrained by the sound speed $c$, 
so the error accompanying this translation is estimated as $\mathcal O (d\partial_n\beta\cdot c\tau_c/N)$.
In the monatomic case, the second term vanishes because of the momentum conservation $\sum_ip_i=0$, then
\begin{eqnarray}
\langle J \rangle^{st}&=&
(\beta_{n+1}-\beta_n)\frac1 N
\int^\infty_0 dt 
\left\langle  
\hat J(0)\hat J(t)\right\rangle^{can-IP}_{(\beta_n,P)}
\nonumber\\&&+\mathcal O\left((d\partial_n\beta)^2,d^2\partial^2_n\beta,d\partial_n\beta \frac{c\tau_c}N\right).
\label{eq:LEGK}
\end{eqnarray}
This is the desired relation. 
Here we defined $\hat J:=\sum_i j_{i+1,i}$.
Apparently, it can be used even if $\Delta \beta$ is comparable with $\bar\beta$.
The perturbation is roughly estimated as $(\Delta\beta d)/(\bar \beta N)$. Smallness of the perturbation comes from the scale separation $d/N\ll1$. 
The reverse transformation of the boundary $\langle\rangle^{can-IP}\to\langle\rangle^{can}$ and the change of the choice of the macroscopic variable $P\to l(T,P)$ recovers the ordinary Green Kubo formula as already discussed in the paper.
\subsection{Remarks}
We note that there is no need to assume the small coupling between the resevoirs and the system if the reservoirs are described as the stochastic ones and satisfy the local detailed balance (\ref{eq:LDB}). It is naturally obeyed by the smallness of the heated area compared to the bulk. So, one can choose any stochastic forces of the thermostats if they keep the local detailed balance. Furthermore, these relations can be applied to the cases of the finite-size systems if it keeps the temperature gaps at the edges negligibly small.

The choice of the reference ensemble determines the accuracy of the expansion, so the reference must be chosen to keep the averages of the macroscopic variables almost unchanged under the ensemble transformation ($P^{st}\to P^{ref}$). Then, the ensemble choice of the homogeneous compression ($l_i=l,P^{ref}=\exp(-\beta_ih_i)/Z$) would not make sense in the far from equilibriums ($\Delta \beta \sim \bar \beta$) in general because such a choice can allow the particles to move in average ($\partial_n P\neq0$). Such a static steady state cannont exsist.
If $\partial P(T,l)/\partial T$ becomes relevant, the difference between two ensembles is not negligible.

As a convincing expample on the validity of the local equilibrium ensembles, 
we show the case of a thermal rectifer~\cite{li2004thermal}.
In that case, the system Hamiltonian takes the form
\begin{eqnarray}
\mathcal H &=& 
\mathcal H_L(x_1,..,x_{N_L},p_1,..,p_{N_L})
\nonumber\\&+&
\mathcal H_R(x_{N_L+1},..,x_N,p_{N_L+1},..,p_N)
\nonumber\\&+&
\lambda V(x_{N_L+1,N_L}) ,\,\,\lambda\ll1. 
\end{eqnarray}
According to the simulation, suppose the situation
\begin{eqnarray}
P=0;\,\, i\leq N_L,T_i=T_L;\,\,i>N_L, T_i=T_R.
\end{eqnarray}
$T_{L(R)}$ is the reservoir temperature of the left (right) side. 
One cannot expand $\Sigma$ by $d\partial_n\beta$ now, but
can get the following relation repeating the same calculation with that of {\it Derivation 2}, defining $q:=j_{N_L+1,N_L}$,
\begin{eqnarray}
\langle q\rangle^{st}
&=&\Delta\beta\int dt
\langle q(0)q(t)\rangle^{lc}_{\beta_L,\beta_R}
+\mathcal O((\Delta\beta)^2\lambda^3)
\hspace{12pt} 
\\&\propto&\lambda^2
\end{eqnarray}
$\Delta \beta:=1/T_R-1/T_L$ is the inverse temperature difference between the left reservoir and the right reservoir.
$\langle\rangle^{lc}_{\beta_L,\beta_R}$ is the average by the local equilibrium ensemble (\ref{eq:lce}) of the corresponding situation.
We expanded $\Sigma$ by the small parameter $\lambda$.
$j_{N_L+1,N_L}\propto \lambda (dV(x_{N_L+1,N_L}))/(dx_{N_L+1,N_L})=\mathcal O(\lambda)$ predicts  the scaling 
$\langle j_{N_L+1,N_L}\rangle^{st}\propto\lambda^2$. This relation is actually observed in~\cite{li2004thermal}.
Also, neglecting the $\mathcal O (\lambda^3)$ terms in the time evolution that come from the system coupling, one can estimate the right hand side with the time-evolution of the separated systems
 ($\mathcal H = \mathcal H_L+\mathcal H_R$), as already done in~\cite{li2004thermal}.
The temperature assymmetry that cannot be treated in the expansions from the global equilibriums is now included in this linear response. 

 We also note the second term in the Green Kubo of the local equilibriums (\ref{eq:LEGK-d}) would remain in the diatomic systems under the compression (tension) even if they recover the normal heat conduction. In addition, even in the case of diatomic systems around the globally equilibriums, the expansion from the isobaric equilibrium shows a different Green Kubo formula which corresponds to (\ref{eq:LEGK-d}) from the iso-dense case (\ref{eq:GEGK}). 
There may be another order because they should be equivalent if we specify the values of all macroscopic variables. Monatomic systems actually have no such confusion. Diatomic systems may need careful modifications on that point.

\section{On the vanishment of \\the temperature gaps at the edges}
In the preceding appendix, we show the simple derivation of the Green Kubo formula and we supposed the 
vanishment of the temperature gaps at the edges (between the system and the reservoirs) there. 
It is reasonable to have doubt whether the gaps really vanish in the simulations.
We showed the data of the temperature profile only at $N=2^{15}=32768$ (FIG \ref{fig:one}), where the gaps still remain.
Here we show the vanishment of the gaps in our simulations of c-FPU-$\beta$.

\begin{figure}[tbp]
  \includegraphics[width=80mm]{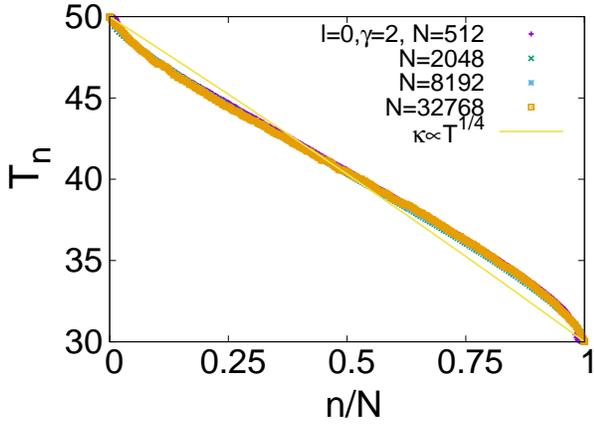}
 \caption{ Temperature profile at c-FPU-$\beta$ at $l=0,\gamma=2,T_L=50,T_R=30 \,\,(T_{L(R)}:$ left (right) reservoir temperature ).
The corresponding curve of normal heat conduction $\kappa\propto T^{1/4}$ is also shown in the same figure. There are the cusps at the edges corresponding to the anomaly but the gaps at the edges cannot be seen.
The clear systematic deviation means the asymptotics of the anomaly realized in this parameter region.
}
 \label{fig:anomaly}
\end{figure}

We begin by seeing that the scale-invariant form of the anomaly can appear at $l=0$ even in $\gamma=2,\Delta T/\bar T=0.5$ ($\Delta T$: temperature difference between two reservoirs).
Figure \ref{fig:anomaly} is the temperature profile at the parameters ($\gamma=2,\Delta T/\bar T=0.5,l=0$).
The $n$-th particle's temperature $T_n$ is defined by the kinetic temperature with the long time average.
One can see the existence of the scale-invariant form
\begin{eqnarray}
T_n=\tilde T_{l=0}(n/N).
\end{eqnarray}
($N$: system size.) As known, The heat conductivity shows power law temperature dependence $\kappa \propto T^{1/4}$ if one takes the temperature difference sufficiently small~\cite{aoki2001fermi} in this parameter $l=0$.
If the system recovers the normal heat condution with the power law temperature dependence of heat conductivity $\kappa\propto T^a$, their corresponding temperature profile must be described as $m\leq n \leq N-m,$ 
\begin{eqnarray}
T_n = \left[ T_{1+m}^{a+1} +(T_{N-m}^{a+1}-T_{1+m}^{a+1})\frac{n-m}{N-2m}
\label{eq:AKR}
\right]^{\frac 1 {a+1}},
\end{eqnarray}
but one cannot see the convergence to it in the figure.
($m$: the number of reservoir-attached particles to each reservoir.) 
There are increase of the gradient (cusps) around the edges (besides the gaps at the edges).
This is a characteristic feature of the anomalous heat conduction and is caused by the growth of the correlation length.
The position-dependent Green Kubo formula (at the position $n$) is proportional to 
$\int dt \sum_i \langle j_{n+1,n}j_{i+1,i}\rangle^{lc}$ ($\langle\rangle^{lc}$:local canonical averange defined in the precedeing appendix), so this is understood as the consequence of the fact that the integration range of the position becomes smaller than the correlation length $(\sim N)$~\cite{van2012exact}.
Here $j_{i+1,i}$ is the energy current between the $i+1,i$-th particles.
Then this systematic deviation corresponds to the anomaly.
Now the temperature gaps at the edges vanish.  Only the cusps remain.

\begin{figure}[t]
  \includegraphics[width=80mm]{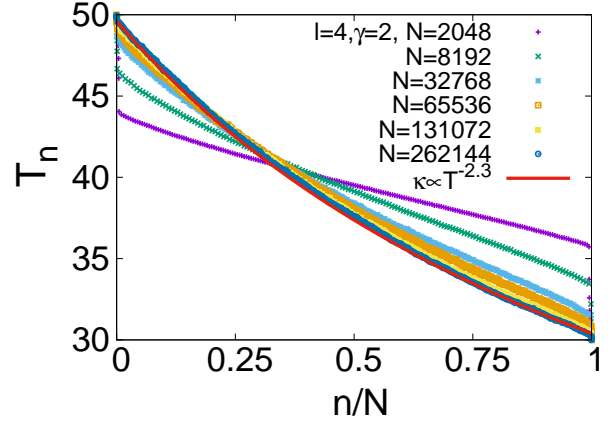}
 \caption{ Temperature profile at c-FPU-$\beta$ at $l=4,\gamma=2,T_L=50,T_R=30$.
The corresponding curve of normal heat conduction $\kappa\propto T^{-2.3}$ is also shown in the same figure. 
The clear convergence means the realization of the asymptotic form in this system.
The gaps at the edges vanish at larger sizes.}
 \label{fig:normal}
\end{figure}

Next, we checked whether the system at $l=4$ reaches the asymptotics (FIG \ref{fig:normal}).
Before the discussion, we note that there are temperature gaps at smaller sizes, but the relation proposed by Aoki \& Kusnezov (\ref{eq:AKR})~\cite{aoki2002nonequilibrium} can be 
used even in the situations.
Then, seeing the profile at small sizes $N=2048,8192,32768$, one can find the increase of the gradient (cusps) at the edges, that corresponds to the anomaly.
It means that there are actually the anomaly in this system at small sizes. 
The graph $\bar \kappa-N$ (FIG \ref{fig:two}) actually has the region that shows the anomaly $\bar \kappa\sim N^{1/3}$ and the region 
just corresponds to the sizes we found here.
After the checking, please see the convergence to the asymptotic curve as the gaps decrease.
The profile that was straight at $N=8192$ has slowly bent and converges to the master curve $\kappa\propto T^{-2.3}$  of normal heat conduction at $N=131072,262144$. 
We put the curve of normal heat conduction $\kappa\propto T^{-2.3}$ with $T_{m+1}=49.7,T_{N-m}=30.3$.
It is clear that it converges to the profile of the normal heat conduction.
At least, one can understand that this behavior is not the artifact that merely cuts the pre-asymptotics only.
The gaps vanish at larger sizes and one can safely use the Green Kubo formula in this case.
\bibliography{reference}

\begin{thebibliography}{68}%
\makeatletter
\providecommand \@ifxundefined [1]{%
 \@ifx{#1\undefined}
}%
\providecommand \@ifnum [1]{%
 \ifnum #1\expandafter \@firstoftwo
 \else \expandafter \@secondoftwo
 \fi
}%
\providecommand \@ifx [1]{%
 \ifx #1\expandafter \@firstoftwo
 \else \expandafter \@secondoftwo
 \fi
}%
\providecommand \natexlab [1]{#1}%
\providecommand \enquote  [1]{``#1''}%
\providecommand \bibnamefont  [1]{#1}%
\providecommand \bibfnamefont [1]{#1}%
\providecommand \citenamefont [1]{#1}%
\providecommand \href@noop [0]{\@secondoftwo}%
\providecommand \href [0]{\begingroup \@sanitize@url \@href}%
\providecommand \@href[1]{\@@startlink{#1}\@@href}%
\providecommand \@@href[1]{\endgroup#1\@@endlink}%
\providecommand \@sanitize@url [0]{\catcode `\\12\catcode `\$12\catcode
  `\&12\catcode `\#12\catcode `\^12\catcode `\_12\catcode `\%12\relax}%
\providecommand \@@startlink[1]{}%
\providecommand \@@endlink[0]{}%
\providecommand \url  [0]{\begingroup\@sanitize@url \@url }%
\providecommand \@url [1]{\endgroup\@href {#1}{\urlprefix }}%
\providecommand \urlprefix  [0]{URL }%
\providecommand \Eprint [0]{\href }%
\providecommand \doibase [0]{http://dx.doi.org/}%
\providecommand \selectlanguage [0]{\@gobble}%
\providecommand \bibinfo  [0]{\@secondoftwo}%
\providecommand \bibfield  [0]{\@secondoftwo}%
\providecommand \translation [1]{[#1]}%
\providecommand \BibitemOpen [0]{}%
\providecommand \bibitemStop [0]{}%
\providecommand \bibitemNoStop [0]{.\EOS\space}%
\providecommand \EOS [0]{\spacefactor3000\relax}%
\providecommand \BibitemShut  [1]{\csname bibitem#1\endcsname}%
\let\auto@bib@innerbib\@empty
\bibitem [{\citenamefont {De~Groot}\ and\ \citenamefont
  {Mazur}(2013)}]{de2013non}%
  \BibitemOpen
  \bibfield  {author} {\bibinfo {author} {\bibfnamefont {S.~R.}\ \bibnamefont
  {De~Groot}}\ and\ \bibinfo {author} {\bibfnamefont {P.}~\bibnamefont
  {Mazur}},\ }\href@noop {} {\emph {\bibinfo {title} {Non-equilibrium
  thermodynamics}}}\ (\bibinfo  {publisher} {Courier Corporation},\ \bibinfo
  {year} {2013})\BibitemShut {NoStop}%
\bibitem [{\citenamefont {Lepri}\ \emph {et~al.}(2003)\citenamefont {Lepri},
  \citenamefont {Livi},\ and\ \citenamefont {Politi}}]{lepri2003thermal}%
  \BibitemOpen
  \bibfield  {author} {\bibinfo {author} {\bibfnamefont {S.}~\bibnamefont
  {Lepri}}, \bibinfo {author} {\bibfnamefont {R.}~\bibnamefont {Livi}}, \ and\
  \bibinfo {author} {\bibfnamefont {A.}~\bibnamefont {Politi}},\ }\href@noop {}
  {\bibfield  {journal} {\bibinfo  {journal} {Physics Reports}\ }\textbf
  {\bibinfo {volume} {377}},\ \bibinfo {pages} {1} (\bibinfo {year}
  {2003})}\BibitemShut {NoStop}%
\bibitem [{\citenamefont {Dhar}(2008)}]{dhar2008heat}%
  \BibitemOpen
  \bibfield  {author} {\bibinfo {author} {\bibfnamefont {A.}~\bibnamefont
  {Dhar}},\ }\href@noop {} {\bibfield  {journal} {\bibinfo  {journal} {Advances
  in Physics}\ }\textbf {\bibinfo {volume} {57}},\ \bibinfo {pages} {457}
  (\bibinfo {year} {2008})}\BibitemShut {NoStop}%
\bibitem [{\citenamefont {Lepri}\ \emph {et~al.}(2015)\citenamefont {Lepri},
  \citenamefont {Livi},\ and\ \citenamefont {Politi}}]{lepri2015heat}%
  \BibitemOpen
  \bibfield  {author} {\bibinfo {author} {\bibfnamefont {S.}~\bibnamefont
  {Lepri}}, \bibinfo {author} {\bibfnamefont {R.}~\bibnamefont {Livi}}, \ and\
  \bibinfo {author} {\bibfnamefont {A.}~\bibnamefont {Politi}},\ }\href@noop {}
  {\bibfield  {journal} {\bibinfo  {journal} {arXiv preprint arXiv:1510.07844}\
  } (\bibinfo {year} {2015})}\BibitemShut {NoStop}%
\bibitem [{\citenamefont {Rieder}\ \emph {et~al.}(1967)\citenamefont {Rieder},
  \citenamefont {Lebowitz},\ and\ \citenamefont {Lieb}}]{rieder1967properties}%
  \BibitemOpen
  \bibfield  {author} {\bibinfo {author} {\bibfnamefont {Z.}~\bibnamefont
  {Rieder}}, \bibinfo {author} {\bibfnamefont {J.}~\bibnamefont {Lebowitz}}, \
  and\ \bibinfo {author} {\bibfnamefont {E.}~\bibnamefont {Lieb}},\ }\href@noop
  {} {\bibfield  {journal} {\bibinfo  {journal} {Journal of Mathematical
  Physics}\ }\textbf {\bibinfo {volume} {8}},\ \bibinfo {pages} {1073}
  (\bibinfo {year} {1967})}\BibitemShut {NoStop}%
\bibitem [{\citenamefont {Casati}\ \emph {et~al.}(1984)\citenamefont {Casati},
  \citenamefont {Ford}, \citenamefont {Vivaldi},\ and\ \citenamefont
  {Visscher}}]{casati1984one}%
  \BibitemOpen
  \bibfield  {author} {\bibinfo {author} {\bibfnamefont {G.}~\bibnamefont
  {Casati}}, \bibinfo {author} {\bibfnamefont {J.}~\bibnamefont {Ford}},
  \bibinfo {author} {\bibfnamefont {F.}~\bibnamefont {Vivaldi}}, \ and\
  \bibinfo {author} {\bibfnamefont {W.~M.}\ \bibnamefont {Visscher}},\
  }\href@noop {} {\bibfield  {journal} {\bibinfo  {journal} {Physical review
  letters}\ }\textbf {\bibinfo {volume} {52}},\ \bibinfo {pages} {1861}
  (\bibinfo {year} {1984})}\BibitemShut {NoStop}%
\bibitem [{\citenamefont {Prosen}\ and\ \citenamefont
  {Robnik}(1992)}]{prosen1992energy}%
  \BibitemOpen
  \bibfield  {author} {\bibinfo {author} {\bibfnamefont {T.}~\bibnamefont
  {Prosen}}\ and\ \bibinfo {author} {\bibfnamefont {M.}~\bibnamefont
  {Robnik}},\ }\href@noop {} {\bibfield  {journal} {\bibinfo  {journal}
  {Journal of Physics A: Mathematical and General}\ }\textbf {\bibinfo {volume}
  {25}},\ \bibinfo {pages} {3449} (\bibinfo {year} {1992})}\BibitemShut
  {NoStop}%
\bibitem [{\citenamefont {Matsuda}\ and\ \citenamefont
  {Ishii}(1970)}]{matsuda1970localization}%
  \BibitemOpen
  \bibfield  {author} {\bibinfo {author} {\bibfnamefont {H.}~\bibnamefont
  {Matsuda}}\ and\ \bibinfo {author} {\bibfnamefont {K.}~\bibnamefont
  {Ishii}},\ }\href@noop {} {\bibfield  {journal} {\bibinfo  {journal}
  {Progress of Theoretical Physics Supplement}\ }\textbf {\bibinfo {volume}
  {45}},\ \bibinfo {pages} {56} (\bibinfo {year} {1970})}\BibitemShut {NoStop}%
\bibitem [{\citenamefont {Rubin}\ and\ \citenamefont
  {Greer}(1971)}]{rubin1971abnormal}%
  \BibitemOpen
  \bibfield  {author} {\bibinfo {author} {\bibfnamefont {R.~J.}\ \bibnamefont
  {Rubin}}\ and\ \bibinfo {author} {\bibfnamefont {W.~L.}\ \bibnamefont
  {Greer}},\ }\href@noop {} {\bibfield  {journal} {\bibinfo  {journal} {Journal
  of Mathematical Physics}\ }\textbf {\bibinfo {volume} {12}},\ \bibinfo
  {pages} {1686} (\bibinfo {year} {1971})}\BibitemShut {NoStop}%
\bibitem [{\citenamefont {O'Connor}\ and\ \citenamefont
  {Lebowitz}(1974)}]{o1974heat}%
  \BibitemOpen
  \bibfield  {author} {\bibinfo {author} {\bibfnamefont {A.}~\bibnamefont
  {O'Connor}}\ and\ \bibinfo {author} {\bibfnamefont {J.}~\bibnamefont
  {Lebowitz}},\ }\href@noop {} {\bibfield  {journal} {\bibinfo  {journal}
  {Journal of Mathematical Physics}\ }\textbf {\bibinfo {volume} {15}},\
  \bibinfo {pages} {692} (\bibinfo {year} {1974})}\BibitemShut {NoStop}%
\bibitem [{\citenamefont {Narayan}\ and\ \citenamefont
  {Ramaswamy}(2002)}]{narayan2002anomalous}%
  \BibitemOpen
  \bibfield  {author} {\bibinfo {author} {\bibfnamefont {O.}~\bibnamefont
  {Narayan}}\ and\ \bibinfo {author} {\bibfnamefont {S.}~\bibnamefont
  {Ramaswamy}},\ }\href@noop {} {\bibfield  {journal} {\bibinfo  {journal}
  {Physical review letters}\ }\textbf {\bibinfo {volume} {89}},\ \bibinfo
  {pages} {200601\_1} (\bibinfo {year} {2002})}\BibitemShut {NoStop}%
\bibitem [{\citenamefont {Dhar}(2001)}]{dhar2001heat}%
  \BibitemOpen
  \bibfield  {author} {\bibinfo {author} {\bibfnamefont {A.}~\bibnamefont
  {Dhar}},\ }\href@noop {} {\bibfield  {journal} {\bibinfo  {journal} {Physical
  review letters}\ }\textbf {\bibinfo {volume} {86}},\ \bibinfo {pages} {5882}
  (\bibinfo {year} {2001})}\BibitemShut {NoStop}%
\bibitem [{\citenamefont {Kundu}\ \emph {et~al.}(2010)\citenamefont {Kundu},
  \citenamefont {Chaudhuri}, \citenamefont {Roy}, \citenamefont {Dhar},
  \citenamefont {Lebowitz},\ and\ \citenamefont {Spohn}}]{kundu2010heat}%
  \BibitemOpen
  \bibfield  {author} {\bibinfo {author} {\bibfnamefont {A.}~\bibnamefont
  {Kundu}}, \bibinfo {author} {\bibfnamefont {A.}~\bibnamefont {Chaudhuri}},
  \bibinfo {author} {\bibfnamefont {D.}~\bibnamefont {Roy}}, \bibinfo {author}
  {\bibfnamefont {A.}~\bibnamefont {Dhar}}, \bibinfo {author} {\bibfnamefont
  {J.~L.}\ \bibnamefont {Lebowitz}}, \ and\ \bibinfo {author} {\bibfnamefont
  {H.}~\bibnamefont {Spohn}},\ }\href@noop {} {\bibfield  {journal} {\bibinfo
  {journal} {EPL (Europhysics Letters)}\ }\textbf {\bibinfo {volume} {90}},\
  \bibinfo {pages} {40001} (\bibinfo {year} {2010})}\BibitemShut {NoStop}%
\bibitem [{\citenamefont {Saito}\ and\ \citenamefont
  {Dhar}(2010)}]{saito2010heat}%
  \BibitemOpen
  \bibfield  {author} {\bibinfo {author} {\bibfnamefont {K.}~\bibnamefont
  {Saito}}\ and\ \bibinfo {author} {\bibfnamefont {A.}~\bibnamefont {Dhar}},\
  }\href@noop {} {\bibfield  {journal} {\bibinfo  {journal} {Physical review
  letters}\ }\textbf {\bibinfo {volume} {104}},\ \bibinfo {pages} {040601}
  (\bibinfo {year} {2010})}\BibitemShut {NoStop}%
\bibitem [{\citenamefont {Spohn}(2015)}]{spohn2015fluctuating}%
  \BibitemOpen
  \bibfield  {author} {\bibinfo {author} {\bibfnamefont {H.}~\bibnamefont
  {Spohn}},\ }\href@noop {} {\bibfield  {journal} {\bibinfo  {journal} {arXiv
  preprint arXiv:1505.05987}\ } (\bibinfo {year} {2015})}\BibitemShut {NoStop}%
\bibitem [{\citenamefont {Lepri}\ \emph {et~al.}(1997)\citenamefont {Lepri},
  \citenamefont {Livi},\ and\ \citenamefont {Politi}}]{lepri1997heat}%
  \BibitemOpen
  \bibfield  {author} {\bibinfo {author} {\bibfnamefont {S.}~\bibnamefont
  {Lepri}}, \bibinfo {author} {\bibfnamefont {R.}~\bibnamefont {Livi}}, \ and\
  \bibinfo {author} {\bibfnamefont {A.}~\bibnamefont {Politi}},\ }\href@noop {}
  {\bibfield  {journal} {\bibinfo  {journal} {Physical review letters}\
  }\textbf {\bibinfo {volume} {78}},\ \bibinfo {pages} {1896} (\bibinfo {year}
  {1997})}\BibitemShut {NoStop}%
\bibitem [{\citenamefont {Hatano}(1999)}]{hatano1999heat}%
  \BibitemOpen
  \bibfield  {author} {\bibinfo {author} {\bibfnamefont {T.}~\bibnamefont
  {Hatano}},\ }\href@noop {} {\bibfield  {journal} {\bibinfo  {journal}
  {Physical Review E}\ }\textbf {\bibinfo {volume} {59}},\ \bibinfo {pages}
  {R1} (\bibinfo {year} {1999})}\BibitemShut {NoStop}%
\bibitem [{\citenamefont {Savin}\ \emph {et~al.}(2002)\citenamefont {Savin},
  \citenamefont {Tsironis},\ and\ \citenamefont {Zolotaryuk}}]{savin2002heat}%
  \BibitemOpen
  \bibfield  {author} {\bibinfo {author} {\bibfnamefont {A.~V.}\ \bibnamefont
  {Savin}}, \bibinfo {author} {\bibfnamefont {G.~P.}\ \bibnamefont {Tsironis}},
  \ and\ \bibinfo {author} {\bibfnamefont {A.~V.}\ \bibnamefont {Zolotaryuk}},\
  }\href@noop {} {\bibfield  {journal} {\bibinfo  {journal} {Physical review
  letters}\ }\textbf {\bibinfo {volume} {88}},\ \bibinfo {pages} {154301}
  (\bibinfo {year} {2002})}\BibitemShut {NoStop}%
\bibitem [{\citenamefont {Mai}\ \emph {et~al.}(2007)\citenamefont {Mai},
  \citenamefont {Dhar},\ and\ \citenamefont {Narayan}}]{mai2007equilibration}%
  \BibitemOpen
  \bibfield  {author} {\bibinfo {author} {\bibfnamefont {T.}~\bibnamefont
  {Mai}}, \bibinfo {author} {\bibfnamefont {A.}~\bibnamefont {Dhar}}, \ and\
  \bibinfo {author} {\bibfnamefont {O.}~\bibnamefont {Narayan}},\ }\href@noop
  {} {\bibfield  {journal} {\bibinfo  {journal} {Physical review letters}\
  }\textbf {\bibinfo {volume} {98}},\ \bibinfo {pages} {184301} (\bibinfo
  {year} {2007})}\BibitemShut {NoStop}%
\bibitem [{\citenamefont {Chang}\ \emph {et~al.}(2008)\citenamefont {Chang},
  \citenamefont {Okawa}, \citenamefont {Garcia}, \citenamefont {Majumdar},\
  and\ \citenamefont {Zettl}}]{chang2008breakdown}%
  \BibitemOpen
  \bibfield  {author} {\bibinfo {author} {\bibfnamefont {C.-W.}\ \bibnamefont
  {Chang}}, \bibinfo {author} {\bibfnamefont {D.}~\bibnamefont {Okawa}},
  \bibinfo {author} {\bibfnamefont {H.}~\bibnamefont {Garcia}}, \bibinfo
  {author} {\bibfnamefont {A.}~\bibnamefont {Majumdar}}, \ and\ \bibinfo
  {author} {\bibfnamefont {A.}~\bibnamefont {Zettl}},\ }\href@noop {}
  {\bibfield  {journal} {\bibinfo  {journal} {Physical review letters}\
  }\textbf {\bibinfo {volume} {101}},\ \bibinfo {pages} {075903} (\bibinfo
  {year} {2008})}\BibitemShut {NoStop}%
\bibitem [{\citenamefont {Landau}\ and\ \citenamefont
  {Lifshitz}(1987)}]{landau1987fluid}%
  \BibitemOpen
  \bibfield  {author} {\bibinfo {author} {\bibfnamefont {L.}~\bibnamefont
  {Landau}}\ and\ \bibinfo {author} {\bibfnamefont {E.}~\bibnamefont
  {Lifshitz}},\ }\href@noop {} {\enquote {\bibinfo {title} {Fluid mechanics (;
  oxford)},}\ } (\bibinfo {year} {1987})\BibitemShut {NoStop}%
\bibitem [{\citenamefont {Grassberger}\ \emph {et~al.}(2002)\citenamefont
  {Grassberger}, \citenamefont {Nadler},\ and\ \citenamefont
  {Yang}}]{grassberger2002heat}%
  \BibitemOpen
  \bibfield  {author} {\bibinfo {author} {\bibfnamefont {P.}~\bibnamefont
  {Grassberger}}, \bibinfo {author} {\bibfnamefont {W.}~\bibnamefont {Nadler}},
  \ and\ \bibinfo {author} {\bibfnamefont {L.}~\bibnamefont {Yang}},\
  }\href@noop {} {\bibfield  {journal} {\bibinfo  {journal} {Physical review
  letters}\ }\textbf {\bibinfo {volume} {89}},\ \bibinfo {pages} {180601}
  (\bibinfo {year} {2002})}\BibitemShut {NoStop}%
\bibitem [{\citenamefont {Mendl}\ and\ \citenamefont
  {Spohn}(2013)}]{mendl2013dynamic}%
  \BibitemOpen
  \bibfield  {author} {\bibinfo {author} {\bibfnamefont {C.~B.}\ \bibnamefont
  {Mendl}}\ and\ \bibinfo {author} {\bibfnamefont {H.}~\bibnamefont {Spohn}},\
  }\href@noop {} {\bibfield  {journal} {\bibinfo  {journal} {Physical review
  letters}\ }\textbf {\bibinfo {volume} {111}},\ \bibinfo {pages} {230601}
  (\bibinfo {year} {2013})}\BibitemShut {NoStop}%
\bibitem [{\citenamefont {Das}\ \emph {et~al.}(2014{\natexlab{a}})\citenamefont
  {Das}, \citenamefont {Dhar}, \citenamefont {Saito}, \citenamefont {Mendl},\
  and\ \citenamefont {Spohn}}]{das2014numerical}%
  \BibitemOpen
  \bibfield  {author} {\bibinfo {author} {\bibfnamefont {S.~G.}\ \bibnamefont
  {Das}}, \bibinfo {author} {\bibfnamefont {A.}~\bibnamefont {Dhar}}, \bibinfo
  {author} {\bibfnamefont {K.}~\bibnamefont {Saito}}, \bibinfo {author}
  {\bibfnamefont {C.~B.}\ \bibnamefont {Mendl}}, \ and\ \bibinfo {author}
  {\bibfnamefont {H.}~\bibnamefont {Spohn}},\ }\href@noop {} {\bibfield
  {journal} {\bibinfo  {journal} {Physical Review E}\ }\textbf {\bibinfo
  {volume} {90}},\ \bibinfo {pages} {012124} (\bibinfo {year}
  {2014}{\natexlab{a}})}\BibitemShut {NoStop}%
\bibitem [{\citenamefont {Kardar}\ \emph {et~al.}(1986)\citenamefont {Kardar},
  \citenamefont {Parisi},\ and\ \citenamefont {Zhang}}]{kardar1986dynamic}%
  \BibitemOpen
  \bibfield  {author} {\bibinfo {author} {\bibfnamefont {M.}~\bibnamefont
  {Kardar}}, \bibinfo {author} {\bibfnamefont {G.}~\bibnamefont {Parisi}}, \
  and\ \bibinfo {author} {\bibfnamefont {Y.-C.}\ \bibnamefont {Zhang}},\
  }\href@noop {} {\bibfield  {journal} {\bibinfo  {journal} {Physical Review
  Letters}\ }\textbf {\bibinfo {volume} {56}},\ \bibinfo {pages} {889}
  (\bibinfo {year} {1986})}\BibitemShut {NoStop}%
\bibitem [{\citenamefont {Mendl}\ and\ \citenamefont
  {Spohn}(2015)}]{mendl2015current}%
  \BibitemOpen
  \bibfield  {author} {\bibinfo {author} {\bibfnamefont {C.~B.}\ \bibnamefont
  {Mendl}}\ and\ \bibinfo {author} {\bibfnamefont {H.}~\bibnamefont {Spohn}},\
  }\href@noop {} {\bibfield  {journal} {\bibinfo  {journal} {Journal of
  Statistical Mechanics: Theory and Experiment}\ }\textbf {\bibinfo {volume}
  {2015}},\ \bibinfo {pages} {P03007} (\bibinfo {year} {2015})}\BibitemShut
  {NoStop}%
\bibitem [{\citenamefont {Kawasaki}\ and\ \citenamefont
  {Oppenheim}(1965)}]{kawasaki1965logarithmic}%
  \BibitemOpen
  \bibfield  {author} {\bibinfo {author} {\bibfnamefont {K.}~\bibnamefont
  {Kawasaki}}\ and\ \bibinfo {author} {\bibfnamefont {I.}~\bibnamefont
  {Oppenheim}},\ }\href@noop {} {\bibfield  {journal} {\bibinfo  {journal}
  {Physical Review}\ }\textbf {\bibinfo {volume} {139}},\ \bibinfo {pages}
  {A1763} (\bibinfo {year} {1965})}\BibitemShut {NoStop}%
\bibitem [{\citenamefont {Pomeau}\ and\ \citenamefont
  {Resibois}(1975)}]{pomeau1975time}%
  \BibitemOpen
  \bibfield  {author} {\bibinfo {author} {\bibfnamefont {Y.}~\bibnamefont
  {Pomeau}}\ and\ \bibinfo {author} {\bibfnamefont {P.}~\bibnamefont
  {Resibois}},\ }\href@noop {} {\bibfield  {journal} {\bibinfo  {journal}
  {Physics Reports}\ }\textbf {\bibinfo {volume} {19}},\ \bibinfo {pages} {63}
  (\bibinfo {year} {1975})}\BibitemShut {NoStop}%
\bibitem [{\citenamefont {Forster}\ \emph {et~al.}(1977)\citenamefont
  {Forster}, \citenamefont {Nelson},\ and\ \citenamefont
  {Stephen}}]{forster1977large}%
  \BibitemOpen
  \bibfield  {author} {\bibinfo {author} {\bibfnamefont {D.}~\bibnamefont
  {Forster}}, \bibinfo {author} {\bibfnamefont {D.~R.}\ \bibnamefont {Nelson}},
  \ and\ \bibinfo {author} {\bibfnamefont {M.~J.}\ \bibnamefont {Stephen}},\
  }\href@noop {} {\bibfield  {journal} {\bibinfo  {journal} {Physical Review
  A}\ }\textbf {\bibinfo {volume} {16}},\ \bibinfo {pages} {732} (\bibinfo
  {year} {1977})}\BibitemShut {NoStop}%
\bibitem [{\citenamefont {Mai}\ and\ \citenamefont
  {Narayan}(2006)}]{mai2006universality}%
  \BibitemOpen
  \bibfield  {author} {\bibinfo {author} {\bibfnamefont {T.}~\bibnamefont
  {Mai}}\ and\ \bibinfo {author} {\bibfnamefont {O.}~\bibnamefont {Narayan}},\
  }\href@noop {} {\bibfield  {journal} {\bibinfo  {journal} {Physical Review
  E}\ }\textbf {\bibinfo {volume} {73}},\ \bibinfo {pages} {061202} (\bibinfo
  {year} {2006})}\BibitemShut {NoStop}%
\bibitem [{\citenamefont {Hu}\ \emph {et~al.}(2000)\citenamefont {Hu},
  \citenamefont {Li},\ and\ \citenamefont {Zhao}}]{hu2000heat}%
  \BibitemOpen
  \bibfield  {author} {\bibinfo {author} {\bibfnamefont {B.}~\bibnamefont
  {Hu}}, \bibinfo {author} {\bibfnamefont {B.}~\bibnamefont {Li}}, \ and\
  \bibinfo {author} {\bibfnamefont {H.}~\bibnamefont {Zhao}},\ }\href@noop {}
  {\bibfield  {journal} {\bibinfo  {journal} {Physical Review E}\ }\textbf
  {\bibinfo {volume} {61}},\ \bibinfo {pages} {3828} (\bibinfo {year}
  {2000})}\BibitemShut {NoStop}%
\bibitem [{\citenamefont {Bricmont}\ and\ \citenamefont
  {Kupiainen}(2007)}]{bricmont2007fourier}%
  \BibitemOpen
  \bibfield  {author} {\bibinfo {author} {\bibfnamefont {J.}~\bibnamefont
  {Bricmont}}\ and\ \bibinfo {author} {\bibfnamefont {A.}~\bibnamefont
  {Kupiainen}},\ }\href@noop {} {\bibfield  {journal} {\bibinfo  {journal}
  {Physical review letters}\ }\textbf {\bibinfo {volume} {98}},\ \bibinfo
  {pages} {214301} (\bibinfo {year} {2007})}\BibitemShut {NoStop}%
\bibitem [{\citenamefont {Pereira}\ and\ \citenamefont
  {Falcao}(2006)}]{pereira2006normal}%
  \BibitemOpen
  \bibfield  {author} {\bibinfo {author} {\bibfnamefont {E.}~\bibnamefont
  {Pereira}}\ and\ \bibinfo {author} {\bibfnamefont {R.}~\bibnamefont
  {Falcao}},\ }\href@noop {} {\bibfield  {journal} {\bibinfo  {journal}
  {Physical review letters}\ }\textbf {\bibinfo {volume} {96}},\ \bibinfo
  {pages} {100601} (\bibinfo {year} {2006})}\BibitemShut {NoStop}%
\bibitem [{\citenamefont {Lepri}\ \emph {et~al.}(1998)\citenamefont {Lepri},
  \citenamefont {Livi},\ and\ \citenamefont {Politi}}]{lepri1998anomalous}%
  \BibitemOpen
  \bibfield  {author} {\bibinfo {author} {\bibfnamefont {S.}~\bibnamefont
  {Lepri}}, \bibinfo {author} {\bibfnamefont {R.}~\bibnamefont {Livi}}, \ and\
  \bibinfo {author} {\bibfnamefont {A.}~\bibnamefont {Politi}},\ }\href@noop {}
  {\bibfield  {journal} {\bibinfo  {journal} {EPL (Europhysics Letters)}\
  }\textbf {\bibinfo {volume} {43}},\ \bibinfo {pages} {271} (\bibinfo {year}
  {1998})}\BibitemShut {NoStop}%
\bibitem [{\citenamefont {Kubo}\ \emph {et~al.}(2012)\citenamefont {Kubo},
  \citenamefont {Toda},\ and\ \citenamefont
  {Hashitsume}}]{kubo2012statistical}%
  \BibitemOpen
  \bibfield  {author} {\bibinfo {author} {\bibfnamefont {R.}~\bibnamefont
  {Kubo}}, \bibinfo {author} {\bibfnamefont {M.}~\bibnamefont {Toda}}, \ and\
  \bibinfo {author} {\bibfnamefont {N.}~\bibnamefont {Hashitsume}},\
  }\href@noop {} {\emph {\bibinfo {title} {Statistical physics II:
  nonequilibrium statistical mechanics}}},\ Vol.~\bibinfo {volume} {31}\
  (\bibinfo  {publisher} {Springer Science \& Business Media},\ \bibinfo {year}
  {2012})\BibitemShut {NoStop}%
\bibitem [{\citenamefont {Gendelman}\ and\ \citenamefont
  {Savin}(2000)}]{gendelman2000normal}%
  \BibitemOpen
  \bibfield  {author} {\bibinfo {author} {\bibfnamefont {O.~V.}\ \bibnamefont
  {Gendelman}}\ and\ \bibinfo {author} {\bibfnamefont {A.~V.}\ \bibnamefont
  {Savin}},\ }\href@noop {} {\bibfield  {journal} {\bibinfo  {journal}
  {Physical review letters}\ }\textbf {\bibinfo {volume} {84}},\ \bibinfo
  {pages} {2381} (\bibinfo {year} {2000})}\BibitemShut {NoStop}%
\bibitem [{\citenamefont {Giardin{\`a}}\ \emph {et~al.}(2000)\citenamefont
  {Giardin{\`a}}, \citenamefont {Livi}, \citenamefont {Politi},\ and\
  \citenamefont {Vassalli}}]{giardina2000finite}%
  \BibitemOpen
  \bibfield  {author} {\bibinfo {author} {\bibfnamefont {C.}~\bibnamefont
  {Giardin{\`a}}}, \bibinfo {author} {\bibfnamefont {R.}~\bibnamefont {Livi}},
  \bibinfo {author} {\bibfnamefont {A.}~\bibnamefont {Politi}}, \ and\ \bibinfo
  {author} {\bibfnamefont {M.}~\bibnamefont {Vassalli}},\ }\href@noop {}
  {\bibfield  {journal} {\bibinfo  {journal} {Physical review letters}\
  }\textbf {\bibinfo {volume} {84}},\ \bibinfo {pages} {2144} (\bibinfo {year}
  {2000})}\BibitemShut {NoStop}%
\bibitem [{\citenamefont {Das}\ and\ \citenamefont {Dhar}(2014)}]{das2014role}%
  \BibitemOpen
  \bibfield  {author} {\bibinfo {author} {\bibfnamefont {S.~G.}\ \bibnamefont
  {Das}}\ and\ \bibinfo {author} {\bibfnamefont {A.}~\bibnamefont {Dhar}},\
  }\href@noop {} {\bibfield  {journal} {\bibinfo  {journal} {arXiv preprint
  arXiv:1411.5247}\ } (\bibinfo {year} {2014})}\BibitemShut {NoStop}%
\bibitem [{\citenamefont {Zhong}\ \emph {et~al.}(2011)\citenamefont {Zhong},
  \citenamefont {Zhang}, \citenamefont {Wang},\ and\ \citenamefont
  {Zhao}}]{zhong2011normal}%
  \BibitemOpen
  \bibfield  {author} {\bibinfo {author} {\bibfnamefont {Y.}~\bibnamefont
  {Zhong}}, \bibinfo {author} {\bibfnamefont {Y.}~\bibnamefont {Zhang}},
  \bibinfo {author} {\bibfnamefont {J.}~\bibnamefont {Wang}}, \ and\ \bibinfo
  {author} {\bibfnamefont {H.}~\bibnamefont {Zhao}},\ }\href@noop {} {\bibfield
   {journal} {\bibinfo  {journal} {arXiv preprint arXiv:1107.3306}\ } (\bibinfo
  {year} {2011})}\BibitemShut {NoStop}%
\bibitem [{\citenamefont {Chen}\ \emph {et~al.}(2012)\citenamefont {Chen},
  \citenamefont {Zhang}, \citenamefont {Wang},\ and\ \citenamefont
  {Zhao}}]{chen2012breakdown}%
  \BibitemOpen
  \bibfield  {author} {\bibinfo {author} {\bibfnamefont {S.}~\bibnamefont
  {Chen}}, \bibinfo {author} {\bibfnamefont {Y.}~\bibnamefont {Zhang}},
  \bibinfo {author} {\bibfnamefont {J.}~\bibnamefont {Wang}}, \ and\ \bibinfo
  {author} {\bibfnamefont {H.}~\bibnamefont {Zhao}},\ }\href@noop {} {\bibfield
   {journal} {\bibinfo  {journal} {arXiv preprint arXiv:1204.5933}\ } (\bibinfo
  {year} {2012})}\BibitemShut {NoStop}%
\bibitem [{\citenamefont {Savin}\ and\ \citenamefont
  {Kosevich}(2014)}]{savin2014thermal}%
  \BibitemOpen
  \bibfield  {author} {\bibinfo {author} {\bibfnamefont {A.~V.}\ \bibnamefont
  {Savin}}\ and\ \bibinfo {author} {\bibfnamefont {Y.~A.}\ \bibnamefont
  {Kosevich}},\ }\href@noop {} {\bibfield  {journal} {\bibinfo  {journal}
  {Physical Review E}\ }\textbf {\bibinfo {volume} {89}},\ \bibinfo {pages}
  {032102} (\bibinfo {year} {2014})}\BibitemShut {NoStop}%
\bibitem [{\citenamefont {Das}\ \emph {et~al.}(2014{\natexlab{b}})\citenamefont
  {Das}, \citenamefont {Dhar},\ and\ \citenamefont {Narayan}}]{das2014heat}%
  \BibitemOpen
  \bibfield  {author} {\bibinfo {author} {\bibfnamefont {S.~G.}\ \bibnamefont
  {Das}}, \bibinfo {author} {\bibfnamefont {A.}~\bibnamefont {Dhar}}, \ and\
  \bibinfo {author} {\bibfnamefont {O.}~\bibnamefont {Narayan}},\ }\href@noop
  {} {\bibfield  {journal} {\bibinfo  {journal} {Journal of Statistical
  Physics}\ }\textbf {\bibinfo {volume} {154}},\ \bibinfo {pages} {204}
  (\bibinfo {year} {2014}{\natexlab{b}})}\BibitemShut {NoStop}%
\bibitem [{\citenamefont {Gendelman}\ and\ \citenamefont
  {Savin}(2014)}]{gendelman2014normal}%
  \BibitemOpen
  \bibfield  {author} {\bibinfo {author} {\bibfnamefont {O.~V.}\ \bibnamefont
  {Gendelman}}\ and\ \bibinfo {author} {\bibfnamefont {A.~V.}\ \bibnamefont
  {Savin}},\ }\href@noop {} {\bibfield  {journal} {\bibinfo  {journal} {EPL
  (Europhysics Letters)}\ }\textbf {\bibinfo {volume} {106}},\ \bibinfo {pages}
  {34004} (\bibinfo {year} {2014})}\BibitemShut {NoStop}%
\bibitem [{\citenamefont {Wang}\ \emph {et~al.}(2013)\citenamefont {Wang},
  \citenamefont {Hu},\ and\ \citenamefont {Li}}]{wang2013validity}%
  \BibitemOpen
  \bibfield  {author} {\bibinfo {author} {\bibfnamefont {L.}~\bibnamefont
  {Wang}}, \bibinfo {author} {\bibfnamefont {B.}~\bibnamefont {Hu}}, \ and\
  \bibinfo {author} {\bibfnamefont {B.}~\bibnamefont {Li}},\ }\href@noop {}
  {\bibfield  {journal} {\bibinfo  {journal} {Physical Review E}\ }\textbf
  {\bibinfo {volume} {88}},\ \bibinfo {pages} {052112} (\bibinfo {year}
  {2013})}\BibitemShut {NoStop}%
\bibitem [{\citenamefont {Di~Cintio}\ \emph {et~al.}(2015)\citenamefont
  {Di~Cintio}, \citenamefont {Livi}, \citenamefont {Bufferand}, \citenamefont
  {Ciraolo}, \citenamefont {Lepri},\ and\ \citenamefont
  {Straka}}]{di2015anomalous}%
  \BibitemOpen
  \bibfield  {author} {\bibinfo {author} {\bibfnamefont {P.}~\bibnamefont
  {Di~Cintio}}, \bibinfo {author} {\bibfnamefont {R.}~\bibnamefont {Livi}},
  \bibinfo {author} {\bibfnamefont {H.}~\bibnamefont {Bufferand}}, \bibinfo
  {author} {\bibfnamefont {G.}~\bibnamefont {Ciraolo}}, \bibinfo {author}
  {\bibfnamefont {S.}~\bibnamefont {Lepri}}, \ and\ \bibinfo {author}
  {\bibfnamefont {M.~J.}\ \bibnamefont {Straka}},\ }\href@noop {} {\bibfield
  {journal} {\bibinfo  {journal} {arXiv preprint arXiv:1509.08796}\ } (\bibinfo
  {year} {2015})}\BibitemShut {NoStop}%
\bibitem [{\citenamefont {Chen}\ \emph {et~al.}(2014)\citenamefont {Chen},
  \citenamefont {Wang}, \citenamefont {Casati},\ and\ \citenamefont
  {Benenti}}]{chen2014nonintegrability}%
  \BibitemOpen
  \bibfield  {author} {\bibinfo {author} {\bibfnamefont {S.}~\bibnamefont
  {Chen}}, \bibinfo {author} {\bibfnamefont {J.}~\bibnamefont {Wang}}, \bibinfo
  {author} {\bibfnamefont {G.}~\bibnamefont {Casati}}, \ and\ \bibinfo {author}
  {\bibfnamefont {G.}~\bibnamefont {Benenti}},\ }\href@noop {} {\bibfield
  {journal} {\bibinfo  {journal} {Physical Review E}\ }\textbf {\bibinfo
  {volume} {90}},\ \bibinfo {pages} {032134} (\bibinfo {year}
  {2014})}\BibitemShut {NoStop}%
\bibitem [{Note1()}]{Note1}%
  \BibitemOpen
  \bibinfo {note} {In the sense that the maximum Lyapunov exponent is large
  compared with dilute case ($d>1$) and the Lyapunov spectra shows the same
  shape with high energy FPU-$\beta $, which correspond to the case that the
  Hessian of Hamiltonian is able to be replaced by the random matrix~\cite
  {newman1986distribution} on the surface spanned by conserved
  quantities.}\BibitemShut {Stop}%
\bibitem [{\citenamefont {Bou-Rabee}\ and\ \citenamefont
  {Owhadi}(2010)}]{bou2010long}%
  \BibitemOpen
  \bibfield  {author} {\bibinfo {author} {\bibfnamefont {N.}~\bibnamefont
  {Bou-Rabee}}\ and\ \bibinfo {author} {\bibfnamefont {H.}~\bibnamefont
  {Owhadi}},\ }\href@noop {} {\bibfield  {journal} {\bibinfo  {journal} {SIAM
  Journal on Numerical Analysis}\ }\textbf {\bibinfo {volume} {48}},\ \bibinfo
  {pages} {278} (\bibinfo {year} {2010})}\BibitemShut {NoStop}%
\bibitem [{Note2()}]{Note2}%
  \BibitemOpen
  \bibinfo {note} {This is a prefetch but one cannot see the gradient
  increasing at the edges in this case, which is a characteristic feature of
  anomalous systems~\cite {van2012exact}.}\BibitemShut {Stop}%
\bibitem [{\citenamefont {van Beijeren}(2012)}]{van2012exact}%
  \BibitemOpen
  \bibfield  {author} {\bibinfo {author} {\bibfnamefont {H.}~\bibnamefont {van
  Beijeren}},\ }\href@noop {} {\bibfield  {journal} {\bibinfo  {journal}
  {Physical review letters}\ }\textbf {\bibinfo {volume} {108}},\ \bibinfo
  {pages} {180601} (\bibinfo {year} {2012})}\BibitemShut {NoStop}%
\bibitem [{\citenamefont {Hurtado}\ and\ \citenamefont
  {Garrido}(2015)}]{hurtado2015violation}%
  \BibitemOpen
  \bibfield  {author} {\bibinfo {author} {\bibfnamefont {P.~I.}\ \bibnamefont
  {Hurtado}}\ and\ \bibinfo {author} {\bibfnamefont {P.~L.}\ \bibnamefont
  {Garrido}},\ }\href@noop {} {\bibfield  {journal} {\bibinfo  {journal} {arXiv
  preprint arXiv:1506.03234}\ } (\bibinfo {year} {2015})}\BibitemShut {NoStop}%
\bibitem [{\citenamefont {Aoki}\ and\ \citenamefont
  {Kusnezov}(2001)}]{aoki2001fermi}%
  \BibitemOpen
  \bibfield  {author} {\bibinfo {author} {\bibfnamefont {K.}~\bibnamefont
  {Aoki}}\ and\ \bibinfo {author} {\bibfnamefont {D.}~\bibnamefont
  {Kusnezov}},\ }\href@noop {} {\bibfield  {journal} {\bibinfo  {journal}
  {Physical review letters}\ }\textbf {\bibinfo {volume} {86}},\ \bibinfo
  {pages} {4029} (\bibinfo {year} {2001})}\BibitemShut {NoStop}%
\bibitem [{\citenamefont {Casati}\ and\ \citenamefont
  {Prosen}(2003)}]{casati2003anomalous}%
  \BibitemOpen
  \bibfield  {author} {\bibinfo {author} {\bibfnamefont {G.}~\bibnamefont
  {Casati}}\ and\ \bibinfo {author} {\bibfnamefont {T.}~\bibnamefont
  {Prosen}},\ }\href@noop {} {\bibfield  {journal} {\bibinfo  {journal}
  {Physical Review E}\ }\textbf {\bibinfo {volume} {67}},\ \bibinfo {pages}
  {015203} (\bibinfo {year} {2003})}\BibitemShut {NoStop}%
\bibitem [{\citenamefont {Seifert}(2012)}]{seifert2012stochastic}%
  \BibitemOpen
  \bibfield  {author} {\bibinfo {author} {\bibfnamefont {U.}~\bibnamefont
  {Seifert}},\ }\href@noop {} {\bibfield  {journal} {\bibinfo  {journal}
  {Reports on Progress in Physics}\ }\textbf {\bibinfo {volume} {75}},\
  \bibinfo {pages} {126001} (\bibinfo {year} {2012})}\BibitemShut {NoStop}%
\bibitem [{\citenamefont {Eyring}(1935)}]{eyring1935activated}%
  \BibitemOpen
  \bibfield  {author} {\bibinfo {author} {\bibfnamefont {H.}~\bibnamefont
  {Eyring}},\ }\href@noop {} {\bibfield  {journal} {\bibinfo  {journal} {The
  Journal of Chemical Physics}\ }\textbf {\bibinfo {volume} {3}},\ \bibinfo
  {pages} {107} (\bibinfo {year} {1935})}\BibitemShut {NoStop}%
\bibitem [{\citenamefont {Spohn}(2014)}]{spohn2014nonlinear}%
  \BibitemOpen
  \bibfield  {author} {\bibinfo {author} {\bibfnamefont {H.}~\bibnamefont
  {Spohn}},\ }\href@noop {} {\bibfield  {journal} {\bibinfo  {journal} {Journal
  of Statistical Physics}\ }\textbf {\bibinfo {volume} {154}},\ \bibinfo
  {pages} {1191} (\bibinfo {year} {2014})}\BibitemShut {NoStop}%
\bibitem [{\citenamefont {Drossel}\ and\ \citenamefont
  {Kardar}(2002)}]{drossel2002passive}%
  \BibitemOpen
  \bibfield  {author} {\bibinfo {author} {\bibfnamefont {B.}~\bibnamefont
  {Drossel}}\ and\ \bibinfo {author} {\bibfnamefont {M.}~\bibnamefont
  {Kardar}},\ }\href@noop {} {\bibfield  {journal} {\bibinfo  {journal}
  {Physical Review B}\ }\textbf {\bibinfo {volume} {66}},\ \bibinfo {pages}
  {195414} (\bibinfo {year} {2002})}\BibitemShut {NoStop}%
\bibitem [{\citenamefont {Lutsko}\ and\ \citenamefont
  {Dufty}(2002)}]{lutsko2002long}%
  \BibitemOpen
  \bibfield  {author} {\bibinfo {author} {\bibfnamefont {J.~F.}\ \bibnamefont
  {Lutsko}}\ and\ \bibinfo {author} {\bibfnamefont {J.~W.}\ \bibnamefont
  {Dufty}},\ }\href@noop {} {\bibfield  {journal} {\bibinfo  {journal}
  {Physical Review E}\ }\textbf {\bibinfo {volume} {66}},\ \bibinfo {pages}
  {041206} (\bibinfo {year} {2002})}\BibitemShut {NoStop}%
\bibitem [{\citenamefont {Kittel}(2005)}]{kittel2005introduction}%
  \BibitemOpen
  \bibfield  {author} {\bibinfo {author} {\bibfnamefont {C.}~\bibnamefont
  {Kittel}},\ }\href@noop {} {\emph {\bibinfo {title} {Introduction to solid
  state physics}}}\ (\bibinfo  {publisher} {Wiley},\ \bibinfo {year}
  {2005})\BibitemShut {NoStop}%
\bibitem [{\citenamefont {Li}\ \emph {et~al.}(2015)\citenamefont {Li},
  \citenamefont {Liu}, \citenamefont {Li}, \citenamefont {H{\"a}nggi},\ and\
  \citenamefont {Li}}]{li20151d}%
  \BibitemOpen
  \bibfield  {author} {\bibinfo {author} {\bibfnamefont {Y.}~\bibnamefont
  {Li}}, \bibinfo {author} {\bibfnamefont {S.}~\bibnamefont {Liu}}, \bibinfo
  {author} {\bibfnamefont {N.}~\bibnamefont {Li}}, \bibinfo {author}
  {\bibfnamefont {P.}~\bibnamefont {H{\"a}nggi}}, \ and\ \bibinfo {author}
  {\bibfnamefont {B.}~\bibnamefont {Li}},\ }\href@noop {} {\bibfield  {journal}
  {\bibinfo  {journal} {New Journal of Physics}\ }\textbf {\bibinfo {volume}
  {17}},\ \bibinfo {pages} {043064} (\bibinfo {year} {2015})}\BibitemShut
  {NoStop}%
\bibitem [{\citenamefont {Kosevich}\ and\ \citenamefont
  {Savin}(2015)}]{kosevich2015confinement}%
  \BibitemOpen
  \bibfield  {author} {\bibinfo {author} {\bibfnamefont {Y.~A.}\ \bibnamefont
  {Kosevich}}\ and\ \bibinfo {author} {\bibfnamefont {A.~V.}\ \bibnamefont
  {Savin}},\ }\href@noop {} {\bibfield  {journal} {\bibinfo  {journal} {arXiv
  preprint arXiv:1509.03219}\ } (\bibinfo {year} {2015})}\BibitemShut {NoStop}%
\bibitem [{\citenamefont {Lepri}\ \emph {et~al.}(2005)\citenamefont {Lepri},
  \citenamefont {Sandri},\ and\ \citenamefont {Politi}}]{lepri2005one}%
  \BibitemOpen
  \bibfield  {author} {\bibinfo {author} {\bibfnamefont {S.}~\bibnamefont
  {Lepri}}, \bibinfo {author} {\bibfnamefont {P.}~\bibnamefont {Sandri}}, \
  and\ \bibinfo {author} {\bibfnamefont {A.}~\bibnamefont {Politi}},\
  }\href@noop {} {\bibfield  {journal} {\bibinfo  {journal} {The European
  Physical Journal B-Condensed Matter and Complex Systems}\ }\textbf {\bibinfo
  {volume} {47}},\ \bibinfo {pages} {549} (\bibinfo {year} {2005})}\BibitemShut
  {NoStop}%
\bibitem [{\citenamefont {Bedeaux}\ and\ \citenamefont
  {Mazur}(1974)}]{bedeaux1974renormalization}%
  \BibitemOpen
  \bibfield  {author} {\bibinfo {author} {\bibfnamefont {D.}~\bibnamefont
  {Bedeaux}}\ and\ \bibinfo {author} {\bibfnamefont {P.}~\bibnamefont
  {Mazur}},\ }\href@noop {} {\bibfield  {journal} {\bibinfo  {journal}
  {Physica}\ }\textbf {\bibinfo {volume} {73}},\ \bibinfo {pages} {431}
  (\bibinfo {year} {1974})}\BibitemShut {NoStop}%
\bibitem [{\citenamefont {Maes}\ and\ \citenamefont
  {Neto{\v{c}}n{\`y}}(2010)}]{maes2010rigorous}%
  \BibitemOpen
  \bibfield  {author} {\bibinfo {author} {\bibfnamefont {C.}~\bibnamefont
  {Maes}}\ and\ \bibinfo {author} {\bibfnamefont {K.}~\bibnamefont
  {Neto{\v{c}}n{\`y}}},\ }\href@noop {} {\bibfield  {journal} {\bibinfo
  {journal} {Journal of Mathematical Physics}\ }\textbf {\bibinfo {volume}
  {51}},\ \bibinfo {pages} {015219} (\bibinfo {year} {2010})}\BibitemShut
  {NoStop}%
\bibitem [{\citenamefont {Itami}\ and\ \citenamefont
  {Sasa}(2015)}]{itami2015nonequilibrium}%
  \BibitemOpen
  \bibfield  {author} {\bibinfo {author} {\bibfnamefont {M.}~\bibnamefont
  {Itami}}\ and\ \bibinfo {author} {\bibfnamefont {S.-i.}\ \bibnamefont
  {Sasa}},\ }\href@noop {} {\bibfield  {journal} {\bibinfo  {journal} {Journal
  of Statistical Physics}\ }\textbf {\bibinfo {volume} {158}},\ \bibinfo
  {pages} {37} (\bibinfo {year} {2015})}\BibitemShut {NoStop}%
\bibitem [{\citenamefont {Li}\ \emph {et~al.}(2004)\citenamefont {Li},
  \citenamefont {Wang},\ and\ \citenamefont {Casati}}]{li2004thermal}%
  \BibitemOpen
  \bibfield  {author} {\bibinfo {author} {\bibfnamefont {B.}~\bibnamefont
  {Li}}, \bibinfo {author} {\bibfnamefont {L.}~\bibnamefont {Wang}}, \ and\
  \bibinfo {author} {\bibfnamefont {G.}~\bibnamefont {Casati}},\ }\href@noop {}
  {\bibfield  {journal} {\bibinfo  {journal} {Physical review letters}\
  }\textbf {\bibinfo {volume} {93}},\ \bibinfo {pages} {184301} (\bibinfo
  {year} {2004})}\BibitemShut {NoStop}%
\bibitem [{\citenamefont {Aoki}\ and\ \citenamefont
  {Kusnezov}(2002)}]{aoki2002nonequilibrium}%
  \BibitemOpen
  \bibfield  {author} {\bibinfo {author} {\bibfnamefont {K.}~\bibnamefont
  {Aoki}}\ and\ \bibinfo {author} {\bibfnamefont {D.}~\bibnamefont
  {Kusnezov}},\ }\href@noop {} {\bibfield  {journal} {\bibinfo  {journal}
  {Annals of Physics}\ }\textbf {\bibinfo {volume} {295}},\ \bibinfo {pages}
  {50} (\bibinfo {year} {2002})}\BibitemShut {NoStop}%
\bibitem [{\citenamefont {Newman}(1986)}]{newman1986distribution}%
  \BibitemOpen
  \bibfield  {author} {\bibinfo {author} {\bibfnamefont {C.~M.}\ \bibnamefont
  {Newman}},\ }\href@noop {} {\bibfield  {journal} {\bibinfo  {journal}
  {Communications in Mathematical Physics}\ }\textbf {\bibinfo {volume}
  {103}},\ \bibinfo {pages} {121} (\bibinfo {year} {1986})}\BibitemShut
  {NoStop}%
\end{thebibliography}%

\end{document}